\documentclass[journal,twoside,web]{ieeecolor}
\usepackage{tmi}
\usepackage{cite}
\usepackage{amsmath,amssymb,amsfonts}
\usepackage{algorithmic}
\usepackage{graphicx}
\usepackage{float}
\usepackage{multirow} 
\usepackage{makecell}
\usepackage{textcomp}
\usepackage{bm}
\usepackage{subfigure}
\usepackage{multirow}
\usepackage{hyperref}
\usepackage{textcomp}
\usepackage{xcolor}
\usepackage{booktabs}
\usepackage{soul}
\usepackage{xcolor}
\newlength{\subfigH}
\usepackage{etoolbox}
\makeatletter
\patchcmd{\@makecaption}
{\scshape}
{}
{}
{}
\makeatletter
\patchcmd{\@makecaption}
{\\}
{.\ }
{}
{}
\makeatother
\def\BibTeX{{\rm B\kern-.05em{\sc i\kern-.025em b}\kern-.08em
    T\kern-.1667em\lower.7ex\hbox{E}\kern-.125emX}}
\begin{document}
\title{Physics Informed Deep Unfolded Full Waveform Inversion for Edema Detection}
\author{Ruizhi Zhang$^{*}$, Yhonatan Kvich$^{*}$,~\IEEEmembership{Graduate Student Member, IEEE}, Rui Guo,~\IEEEmembership{Member, IEEE}, \\ 
	Oded Cohen, Yonina C. Eldar,~\IEEEmembership{Fellow,~IEEE}
	% <-this % stops a space
	\thanks{* Equal contribution.
%	}
%	\thanks{
	Ruizhi Zhang is with University of Electronic Science and Technology of China, China, and also with Faculty of Math and Computer Science, Weizmann Institute of Science, Israel. Yhonatan Kvich, Rui Guo, Oded Cohen and Yonina C. Eldar are with Faculty of Math and Computer Science, Weizmann Institute of Science, Israel. 
%		Ruizhi Zhang is also with University of Electronic Science and Technology of China, China.
	}
	\thanks{This research was supported by the Israel Science Foundation (grant No. 3805/21), within the Israel Precision Medicine Partnership (IPMP) program, by the European Research Council (ERC) under the European Union's Horizon 2020 research and innovation program (grant No. 101000967), as well as by Israel Science Foundation (grant No. 536/22).}% <-this % stops a space
	%\thanks{Manuscript received April 19, 2021; revised August 16, 2021.}
}

\IEEEaftertitletext{\vspace{-1.6em}}
\maketitle

\begin{abstract}
Edema is a potential indicator of underlying pathological changes. However, its low-contrast signature is often masked in conventional B-mode imaging by strong scatterers, making reliable detection challenging.
Ultrasound (US) provides a non-invasive, non-ionizing, and cost-efficient imaging option that is widely used. Conventional techniques, which rely on beamforming, often lack sufficient physical interpretability. Quantitative US (QUS) can estimate physical properties such as the speed of sound (SoS) and density by solving a physics-based inverse problem directly on the measured US wavefields, i.e., the raw per-element channel data (CD), to recover their spatial distribution.
However, state-of-the-art physics-based inversion methods, including full waveform inversion (FWI) and model-based quantitative radar and US (MB-QRUS), are computationally intensive and susceptible to local minima, which constrains their clinical utility. 
We introduce deep unfolded FWI (DUFWI), a physics-faithful unfolded iterative inversion method that exhibits FWI-like refinement behavior while learning the update rule from data, requiring only a small number of iterations for real-time SoS reconstruction.
Across both simulated datasets and hardware measurements acquired with a Verasonics US system, the DUFWI significantly outperforms classical FWI and MB-QRUS in reconstruction quality while maintaining high computational efficiency. These results demonstrate real-time edema diagnosis in both simulation and hardware experiments, with phantom-based validation using cylindrical rods, supporting practical deployment under typical US imaging setting.
\vspace{-0.576cm}
\end{abstract}
\begin{IEEEkeywords}
		ultrasound, full waveform inversion, deep unfolding, medical imaging, model-based networks.
\end{IEEEkeywords}

\vspace{-0.2cm}
\section{Introduction}
\vspace{-0.13cm}
\label{sec:introduction}
Arm (upper-limb) edema is the abnormal accumulation of fluid within arm tissue. Timely detection and grading of disease severity guide treatment and enable longitudinal monitoring~\cite{bn-1}. Ultrasound (US) is well suited to arm-edema diagnosis and follow-up because it is non-invasive, non-ionizing, portable, and cost-efficient, allowing bedside and repeated assessments~\cite{bn-2,bn1}. US images are formed from channel data (CD), the raw echo signals recorded on multiple receives. These echoes are produced by the medium in response to sequences of acoustic pulses transmitted from multiple transducer elements~\cite{bn2}.

Conventional B-mode edema assessment relies on beamforming (BF), where received signals from multiple transducer elements are time-aligned and coherently summed with appropriate apodization to enhance echoes from focal regions while suppressing off-axis noise~\cite{bn3}. While effective for structural visualization, BF does not explicitly model tissue acoustic properties that are directly affected by fluid content, such as speed of sound (SoS) and density. As a result, US-based tissue assessment remains limited in specificity and quantitative comparability across patients and time, including for edema~\cite{bn-3}.

Quantitative US (QUS) addresses these limitations by estimating tissue-specific acoustic parameters including SoS and density~\cite{bn5}. These parameters are mechanistically linked to water content and microstructure, providing biomarkers for more sensitive edema detection, objective grading, and standardized longitudinal tracking. A QUS-based assessment of limb lymphedema is introduced in~\cite{edema1}, by using subcutaneous thickness and shear-wave-derived stiffness to distinguish the affected limb from the contralateral limb. 
%Subsequent work~\cite{edema2} showed that the same QUS metrics can be followed over time to monitor response to therapy. 
Beyond edema, QUS has demonstrated utility in cancer detection, tissue characterization, bone analysis, and rapid stroke imaging~\cite{bn6,bn7,bone1}.

In a widely used QUS approach, tissue property estimation is formulated as an inverse problem, in which a physically forward model of acoustic wave propagation links spatially varying parameters to the CD. Given CD, spatial parameter maps are estimated by minimizing a regularized data misfit between measured and simulated responses, which yields a nonlinear inverse problem~\cite{bn8}. 
Full-waveform inversion (FWI) employs gradient-based optimization to reconstruct properties such as SoS and density,
which originated from geophysics and then adpated to medical US~\cite{bn9}. Despite its promise, FWI remains computationally demanding and is fundamentally limited by the severe nonlinearity and ill-posedness of the underlying inverse problem, which slows reconstruction and risk convergence to local minima. These challenges limit the application of FWI in medical imaging.
%, where precise and timely reconstructions are crucial. 

Recently, deep learning methods have been proposed to solve the nonlinear inverse problem in US imaging, enabling real-time reconstructions and reducing the risk of convergence to local minima~\cite{bn10}. 
By incorporating the wave-equation forward operator, these networks are guided by the explicit physics model,
yielding more accurate reconstructions with fewer trainable parameters~\cite{bn13,bn14}. 
%For instance, in \cite{bn15}, each finite-difference time-domain (FDTD) time-step is implemented as a recurrent neural cell whose convolutional kernels enforce the acoustic wave equation, providing an end-to-end differentiable time-domain simulator that enables SoS reconstruction by back-propagating the data misfit.
For instance, in \cite{bn15}, each finite-difference time-domain (FDTD) time-step is implemented as a recurrent neural cell whose convolutional kernels enforce the acoustic wave equation, enabling SoS reconstruction by back-propagating the data misfit.
However, such methods may exhibit training instability due to the dependence on partial differential equations during backpropagation and demand fine discretizations. In addition, several model-driven deep learning frameworks have been proposed~\cite{bn18,bn19,bn20}, but they often exhibit degraded convergence under discretization constraints and cannot fully utilize the domain knowledge.

Alternatively, deep unfolding (DU) has recently attracted considerable attention as a powerful tool for solving inverse problems~\cite{bn21,bn22,bn23,bn24}. By unfolding a conventional iterative algorithm into a finite sequence of trainable layers, DU seamlessly integrates physics-based modeling with data-driven learning~\cite{bn25,bn26,bn27,bn28}. Examples include the learned primal-dual framework for tomographic reconstruction~\cite{bn29} and unfolding for general imaging tasks~\cite{bn30}. Building on these advances, physics-informed DU has been proposed that embeds physical constraints directly into the network architecture~\cite{bn31,bn32}. In the context of US imaging, DU methods have likewise shown great promise for quantitative reconstruction. Notably, the model-based quantitative radar and US (MB-QRUS) approach in~\cite{MBQRUS} operates directly on raw CD and embeds the explicit acoustic forward model into a network, using a learned gradient-descent update within a single unfolded iteration to recover physical property maps. However, a key conceptual advantage of unfolding iterative physics-based inversion is the ability to exhibit iterative refinement.
Although MB‑QRUS offers high computational efficiency, its reliance on a single iteration may limit its ability to capture fine structural details.

Motivated by the work in microwave imaging~\cite{bn34}, we propose a deep unfolded FWI (DUFWI) framework tailored to inverse US, with the goal of preserving the FWI-like iterative correction while learning the update rule, enabling real-time reconstruction with only a few iterations.
In our approach, each unfolded iteration uses a neural update module that takes the data-fidelity gradient, computed from full-bandwidth time-domain waveforms, to form an update. This methodology maintains a strong link to the physical knowledge while allowing the update rule to be learned from data by the network. For training efficiency, we adopt a lightweight block-wise training strategy that avoids backpropagating through the full unfolding iteration chain and the associated forward operators. Instead, each iteration is trained as a local subproblem, which substantially reduces both training memory consumption and runtime. 
After training, the network predicts the SoS update at each iteration based on the data-fidelity gradient. 

We evaluate our method on three types of data:
(i) SoS maps that are based on MNIST shapes, where the SoS values were chosen to match those observed in real tissues, (ii) clinically motivated simulations of a human arm with or without edema, and (iii) hardware measurements on a tissue-mimicking phantom. 
In numerical and hardware experiments, DUFWI achieves better performance than classical FWI and MB-QRUS, and requires much fewer iterations than classical FWI.
These results highlight the efficiency of our approach, making it a promising solution for practical real-time QUS applications. While this paper focuses on edema detection, we emphasize that the underlying inverse task is general and can yield rich quantitative information across diverse applications.

Preliminary work and results were presented in~\cite{icassp_cof}. Herein, we present the full derivation of DUFWI and examine the method under diverse testing condition, and provide a detailed analysis of its advantages. In addition, we design a dedicated measurement system, acquire real CD from a tissue-mimicking phantom, and apply existing inverse methods to these hardware measurements to validate the method beyond simulation. We further describe the implementation details, the clinically motivated dataset, and the hardware evaluation results.

%This paper is organized as follows: Section \ref{sec: problem_statement} formulates the forward model and the associated inverse problem, and reviews conventional FWI and MB-QRUS.
The remainder of this paper is organized as follows: Section \ref{sec: problem_statement} describes the forward model and the associated inverse problem, and reviews conventional FWI and MB-QRUS.
Section \ref{sec:inversion_methods} presents our DU approach to FWI. Section \ref{sec:harware system} details our designed hardware system used to evaluate performance under realistic conditions. Numerical and hardware results are shown in Section \ref{subsec: results} and Section \ref{hardware res}, respectively. Section \ref{sec: conclusion} offers concluding insights and discusses potential future directions.

\vspace{-0.2cm}
\section{Foundations and Problem Statement}\label{sec: problem_statement}  
%\vspace{-0.2cm}
\subsection{Forward Modeling}\label{subsec:Problem Formulation}
\vspace{-0.06cm}
\begin{figure}[t!] 
	\vspace*{-0.5em}
	\centering
	{{\includegraphics[width=0.406\textwidth]{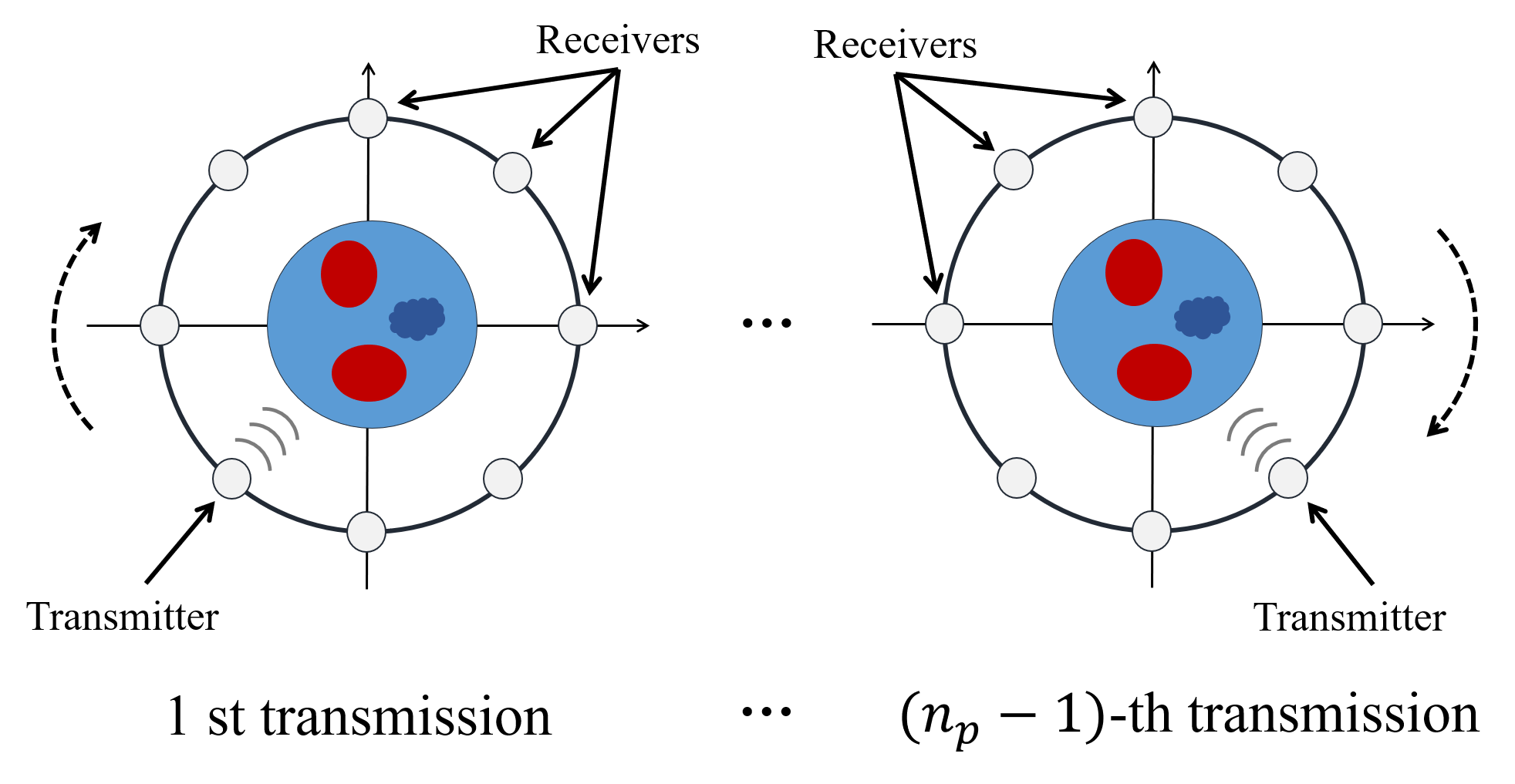}}}
	\vspace*{-0.86em}
	\caption{A typical 2-D measurement setup for edema diagnosis, illustrating a target arm surrounded by a circular array of transducers.}  
	\label{fig1} 
	\vspace*{-1.83em}
\end{figure}
In this work, we consider an US imaging system with a circular transducer array of $n_p$ transducer elements. Each element can act as both a transmitter and a receiver. We introduce a sequential transmission scheme, in which a single transducer acts as the transmitter while a subset of other elements of size $n_R$ serves as receivers ($n_R\leq n_p$); the receiver subset can change between transmissions. Each element transmits once, yielding $n_p$ transmission in total and $n_p \times n_R$ received signals. 
%During each transmission, a single transducer acts as the transmitter, while $n_R$ other transducers serve as receivers. Each transducer takes turns as the transmitter, resulting in a total of $n_p$ transmission totally and $n_p \times n_R$ received signal. 
This configuration allows for comprehensive spatial sampling and improved resolution through the integration of multiple independent measurements. The goal of this paper focuses on reconstruction of SoS within the scanned medium, which provides clinical information. A typical 2-D setup for US imaging is illustrated in Fig.~\ref{fig1}.

%The scattering field, denoted by $u(t,x,z)$, changes in space and time and is related to the physical properties of the scanned medium by the wave propagation equation.
%For US, the wave propagation equation is expressed as~\cite{Absorb}
The wave propagation equation relates the physical properties of the scanned medium to the scattering field $u(t,x,z)$, which varies in both space and time. In US scenario, the equation is written as~\cite{Absorb}
\begin{align}\label{eq:wave prop}
	c_0^2 & \rho_0 \left( \frac{\partial}{\partial x} \left( \frac{1}{\rho_0} \frac{\partial u}{\partial x} \right) + \frac{\partial}{\partial z} \left( \frac{1}{\rho_0} \frac{\partial u}{\partial z} \right) \right) + S \nonumber\\
	&= \frac{\partial^2 u}{\partial t^2} + 2D \frac{\partial u}{\partial t} + D^2 u,
\end{align}
where $c_0(x,z)$ denotes the SoS of the medium and serves as the target parameter in the reconstruction, $\rho_0(x,z)$ corresponds to the medium density, $S(t,x,z)$ is the source pulse, and $D(x,z)$ represents the artificial damping profile used to suppress boundary reflections and allow a finite computational domain. 
We implement $D(x,z)$ as perfectly matched layers (PMLs) following the absorbing-boundary formulation in \cite{PML,Absorb}, which provides efficient and robust attenuation near the borders. 
For brevity, in equation \eqref{eq:wave prop} we omitted the brackets for $S, c_0, D, \rho_0$. We assume that the medium density is constant and set to $\rho_0 = 1000$ $kg/m^3$~\cite{MBQRUS}.

%To obtain the discrete wave propagation equations, we adopt a grid of size $n_x \times n_z$ and discretize the temporal and spatial derivatives\cite{bn8,Absorb}. The temporal derivative can be obtained by a weighted combination of previous time-step values, whereas the spatial derivative is given by a convolution with a Laplacian or gradient kernel. 
Adopting a grid of size $n_x \times n_z$, we obtain the discrete wave propagation equations by discretizing the temporal and spatial derivatives \cite{bn8,Absorb}, where the former is computed as a weighted combination of previous temporal samples and the latter is computed through convolution with a Laplacian or gradient kernel.
We define the discrete scattering field and source pulse as $\textbf{U}, \textbf{S} \in \mathbb{R}^{n_x\times n_z \times n_t}$, respectively, where $n_t$ denotes the number of discrete time samples. 
%Furthermore, $\textbf{U}[t]\in\mathbb{R}^{n_x \times n_z}$ denotes the discrete scattering field at the $t$-th time step (similarly $\textbf{S}[t]$). 
Moreover, the discrete scattering field at the $t$-th time step is denoted by $\textbf{U}[t]\in\mathbb{R}^{n_x \times n_z}$, and $\textbf{S}[t]$ is defined similarly.
After organizing the discrete wave propagation equation, $\textbf{U}[t]$ can be written as
%To obtain discrete wave propagation equations, we use a discrete grid with size $n_x \times n_z$ and a discrete form of the time and spatial derivatives \cite{bn8,Absorb}. The discrete-time derivative is given by a weighted average of the past time samples, and the discrete-spatial derivative is given by a convolution with the Laplacian or gradient kernel. We denote $\textbf{U}, \textbf{S} \in \mathbb{R}^{n_x\times n_z \times n_t}$ as the discrete scattering field and source pulse, respectively, where $n_t$ is the number of discrete-times samples. In addition, we denote $\textbf{U}[t]\in\mathbb{R}^{n_x \times n_z}$ as the discrete scattering field for the $t$-th time step (similarly $\textbf{S}[t]$). After organizing the discrete US wave propagation equation, $\textbf{U}[t]$ can be written as
\begin{align}\label{eq:initial forward}
	\textbf{U}[t] = &\frac{\textbf{2} - \textbf{D}^2 \Delta_t^2}{\textbf{1}+\textbf{D}\Delta_t}\textbf{U}[t-1] + \frac{\textbf{D}\Delta_t - \textbf{1}}{\textbf{1} + \textbf{D}\Delta_t} \odot \textbf{U}[t-2]\nonumber \\
	& + \frac{\textbf{C}^2\Delta_t }{\textbf{1}+\textbf{D}\Delta_t}\odot (\nabla_D^2 * \textbf{U}[t-1]) + \textbf{S}[t].
\end{align}
%Here, $\odot$ is element-wise multiplication, $*$ is the convolution operator, $\nabla_D$ is the discrete gradient filter, $\textbf{1}\in\mathbb{R}^{n_x \times n_z}$ is a matrix of all ones, $\textbf{2}\in\mathbb{R}^{n_x \times n_z}$ is a matrix of all twos, $\Delta_t\in\mathbb{R}^{n_x \times n_z}$ is a matrix with the value $dt$ for each entry, $\textbf{C}, \textbf{D} \in \mathbb{R}^{n_x \times n_z}$ are the discrete SoS, and damping, respectively.
Here, $\odot$ and $*$ denote element-wise multiplication and convolution, respectively. The symbol $\nabla_D$ represents the discrete gradient filter. $1 \in \mathbb{R}^{n_x \times n_z}$ and $2 \in \mathbb{R}^{n_x \times n_z}$ denote matrices whose entries are all one and two, respectively, $\Delta_t \in \mathbb{R}^{n_x \times n_z}$ is a matrix whose entries are all $dt$, and $C, D \in \mathbb{R}^{n_x \times n_z}$ represent the discrete SoS and damping, respectively.

The above forward model describes the transmission of the waveform to the receivers under given physical properties. The pressure field is recorded at the receiver locations.
%We denote the measured CD for the $p$-th transmission as $\mathbf{M}[p]\in\mathbb{R}^{n_t \times n_R}$ which consists of $n_t$ time samples and $n_R$ receiving channels. 
Let $\mathbf{M}[p]\in\mathbb{R}^{n_t \times n_R}$ denote the measured CD obtained from the $p$-th transmission, where $n_t$ and $n_R$ represent the numbers of time samples and receiving channels, respectively.
For notational convenience, the forward model for the \(p\)-th transmission is written as
\begin{equation}\label{eq:forward}
	\mathbf{M}[p] = \tilde{F}_{p}(\mathbf{C}),
\end{equation}
where \(\tilde{F}_{p}\) maps the SoS field \(\mathbf{C}\) to \(\mathbf{M}[p]\) by solving the wave equation and applying the receiver/time-sampling operator. Collecting all transmissions yields
\begin{equation}\label{eq:forward_all}
	\mathbf{M} = \tilde{F}(\mathbf{C}),
\end{equation}
where $\tilde{F}(\mathbf{C}) \triangleq \bigl\{\tilde{F}_{p}(\mathbf{C})\bigr\}_{p=1}^{n_p}$. Our goal is to reconstruct $\textbf{C}$ from $\mathbf{M}$, which yields a nonlinear, non-convex, and ill-posed inverse problem in heterogeneous tissues. The following subsections discuss two approaches to solve the inverse problem.

\vspace{-0.5cm}
\subsection{Conventional Approach} \label{subsec:conventional approach}
\vspace{-0.1cm}

\begin{figure}[t!]
	\centering
	\vspace*{-0.06em}
	\includegraphics[width=0.426\textwidth, keepaspectratio]{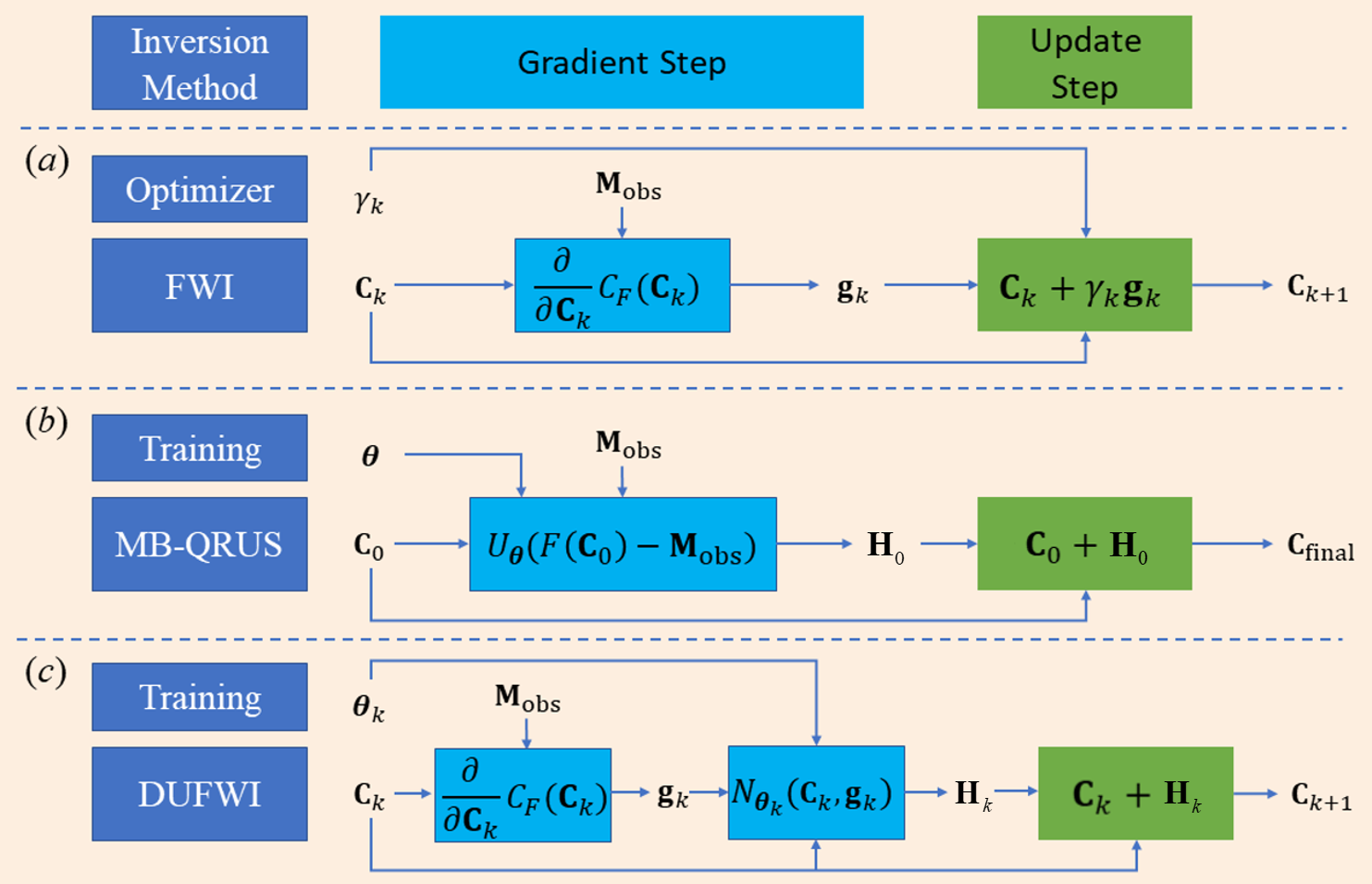}
	\vspace*{-0.39em}
	\caption{Visual comparision of the gradient update of three inverse methods: (a) FWI, (b) MB-QRUS, (c) DUFWI.}
	\label{fig: classical}
	\vspace*{-2.0em}
\end{figure}

\subsubsection{FWI} \label{subsubsec:fwi}
%As a physics-based approach, FWI directly minimizes the discrepancy iteratively between the observed data \(\mathbf{M}_\text{obs}\) and the simulated data \(\tilde{F}(\mathbf C)\)
%, while SoS map is the optimization variables~\cite{bn9}. 
%This is typically formulated as the minimization of the objective
%\begin{equation} \label{eq: cost_SoS}
%	C_F:=\| \mathbf{M}_{\text{obs}} - \tilde{F}(\mathbf C)\|_2^2 + \lambda R(\mathbf C),
%\end{equation}
As a physics-based approach, FWI seeks to estimate $\mathbf C$ by iteratively minimizing the discrepancy between the observed CD $\mathbf{M}_\text{obs}$ and $\tilde{F}(\mathbf C)$~\cite{bn9}, given by
\begin{equation}\label{eq:cost_SoS}
	C_F := \|\mathbf{M}_\text{obs} - \tilde{F}(\mathbf C)\|_2^2 + \lambda R(\mathbf C).
\end{equation}
where the first term $\| \mathbf{M}_{\text{obs}} - \tilde{F}(\mathbf C)\|_2^2$ is the data fidelity term that measures the mismatch between the observed data and the simulated measurements, and $\lambda R(\mathbf C)$ is the regularization term that imposes prior constraints to the solution space~\cite{Rudin1992}. In many practical settings, $R(\mathbf C)$ is chosen to be non-smooth (e.g., total variation or other $\ell_1$-type penalties), so the overall optimization problem can be handled by operator-splitting schemes such as ADMM, where the regularization is enforced via proximal updates rather than explicit gradients~\cite{admm}.
%$R$ denotes the regularization function, which typically incorporates constraints such as smoothness or total variation (TV)~\cite{Rudin1992}. 
Within this framework, the main computational cost lies in computing the gradient of the data fidelity term with respect to the SoS map.
%	the main challenge is the computation of the gradient of the data-fidelity term with respect to the SoS map. 
%	In our DUFWI formulation, we therefore focus on evaluating the gradient of the data misfit, while the effect of the regularization is incorporated at the level of the optimization algorithm. 
Given a current SoS estimate at iteration $k$, the gradient of the data-fidelity term is denoted as
\begin{equation}\label{eq: get_grad}
	\mathbf{g}_k:= \frac{\partial}{\partial \mathbf{C}_{k}}\| \mathbf{M}_{\text{obs}} - \tilde{F}(\mathbf C_{k})\|_2^2.
\end{equation}
%Despite their widespread adoption, these handcrafted regularizers are often insufficient to characterize the necessary features for accurate reconstruction, as we demonstrate in the numerical examples.
%The minimization of (\ref{eq: cost_SoS}) is typically performed using gradient-based methods, such as conjugate gradient or L-BFGS \cite{bn8}. These approaches iteratively update the SoS estimate by computing the gradient of the cost function with respect to the current SoS estimation. Given a SoS map at iteration $k$, the gradient of the data fidelity function is denoted as
%\begin{equation}\label{eq: get_grad}
%	\mathbf{g}_{k} = \frac{\partial}{\partial \mathbf{C}_{k}} C_{F}(\mathbf{C}_{k}).
%\end{equation}
Based on the computed gradient at the $k$-th iteration $\mathbf{g}_{k}$, the SoS map is iteratively updated by
\begin{equation} \label{eq: SoS_update}
	\mathbf C_{k+1} = \mathbf C_k + \gamma_k \mathbf g_k, 
\end{equation}
where \(\gamma_k\) is the step size at the \(k\)-th iteration. See Fig.~\ref{fig: classical}(a) for an outline of the FWI method.

Nevertheless, the optimization problem underlying FWI is not only highly non-convex but also computationally intensive due to repeated forward wavefield simulation and the subsequent gradient computation.
%, which poses significant challenges for conventional gradient-based methods. 
These algorithms rely solely on local gradient information and are therefore susceptible to being trapped in suboptimal solutions. Due to the strong non-convexity of the problem, convergence often requires hundreds or even thousands of iterations, with each involving both forward wave propagation and gradient computation, making the process time-consuming. 
Reconstructing a single image may take several hours or even days, and often fails to meet the accuracy demands of practical applications.

\subsubsection{MB-QRUS} \label{subsec: MB-QRUS}

%To overcome the heavy computation and sensitivity to local minima of FWI, MB-QRUS is leveraged \cite{MBQRUS} as a model-driven deep-learning architecture, combining physics-guided modeling with adaptive learning to achieve faster and accurate reconstructions. 
To address FWI's high computational burden and susceptibility to local minima, MB-QRUS was introudced in~\cite{MBQRUS} and integrates physics-guided modeling with adaptive learning for faster and more accurate reconstruction.
MB-QRUS performs a single update, defined as
\begin{equation} \label{eq: SoS_update_deep}
	\mathbf{C}_{\rm final} = \mathbf{C}_{0} +  \mathbf{H}_{0},
\end{equation}
where \(\mathbf{C}_0\) corresponds to a uniform SoS map, equal to the value of the background, \(\mathbf{C}_{\rm final}\) is the final estimated SoS map, and \(\mathbf{H}_0 \in \mathbb{R}^{n_x \times n_z} \) is the update term. This update is predicted by a U-Net-based network $U_{\theta}$, which adaptively determines both the update direction and the step size.
%The network incorporates in its design the physical model of wave propagation \eqref{eq:initial forward} to calculate $\tilde{F}(\textbf{C}_0) \in\mathbb{R}^{n_t \times n_p \times n_R}$ for $\textbf{C}_0$, 
The network is designed to incorporate the physical wave propagation model in \eqref{eq:initial forward} to compute $\tilde{F}(\textbf{C}_0) \in\mathbb{R}^{n_t \times n_p \times n_R}$ for $\textbf{C}_0$,
and uses the data residual $\tilde{\textbf{M}} = \mathbf{M}_{\text{obs}} - \tilde{F}(\mathbf C_0)$ as input to $U_{\theta}$
to predict $\textbf{H}_0$. In addition, $\mathbf C_0$ are also given as input to $U_{\theta}$.
A visual schematic of MB-QRUS is provided in Fig.~\ref{fig: classical}(b).

By embedding physical priors directly into the network design, it strikes a balance between interpretability and learning flexibility. In addition, MB-QRUS operates directly in the CD domain, without requiring explicit transformations into the image domain. However, it may underuse available physical and model-based information due to limited incorporation of physics models, which limit reconstruction quality in highly complex scenarios. 
%Without iterative refinement, the model cannot capture the optimization dynamics that are central to classical FWI, where gradual updates drive accuracy and convergence. 
%Without iterative refinement, the model cannot reproduce the iterative update behavior that is central to FWI, in which successive gradient-based updates progressively reduce the data misfit and drive convergence toward the true SoS map.
Without iterative refinement, the model cannot emulate the core optimization mechanism of FWI and is confined to a single-step estimate, rather than the progressive reduction of data misfit achieved through successive gradient-based updates. As a result, generalization across diverse initializations and heterogeneous tissues can be challenging.

%	Nevertheless, as it approximates only a single iteration of the underlying optimization process, it may underutilize the full range of physical and model-based information, potentially limiting its reconstruction capability in highly complex scenarios.
%	While MB-QRUS demonstrates promising results by using a learned model to approximate the gradient descent from an initial speed of sound map $\mathbf{C}_0$ to a final reconstruction $\mathbf{C}_{\text{final}}$, it effectively simulates only a single-step update. 
%	Such a one-shot approach limits the ability of model to capture the iterative nature of traditional FWI, where gradual refinement across multiple updates is critical for accuracy and convergence. In other words, compressing the entire inversion process into one large step makes it difficult for the network to generalize across diverse initializations or complicated tissue structures. 

\vspace{-0.3cm}
\section{Deep Unfolding of Full Waveform Inversion} \label{sec:inversion_methods}
\vspace{-0.05cm}
\subsection{DUFWI Framework}
\vspace{-0.1cm}
To address the limitations of conventional FWI and single-iteration MB-QRUS,
we propose a deep unfolding FWI algorithm, which decomposes the inversion into multiple learnable update steps. 
Each stage is explicitly modeled by a neural network, which takes as input the SoS estimate from the previous stage along with the corresponding gradient of the data fidelity~\eqref{eq: get_grad}, and produces an updated SoS prediction. 
These networks are trained sequentially to progressively refine the SoS estimate, leading to improved training stability.

Let $\mathbf{C}_{k}$ and $\mathbf{C}_{k+1}$ denote the SoS map estimates at the $k$-th and $(k+1)$-th iterations, respectively. The update at the $k$-th iteration is defined as
\begin{equation}\label{eq:dufwi_update}    
	\mathbf{C}_{k+1} = \mathbf{C}_{k} + \mathbf{H}_{k},
\end{equation}
where the update term $\mathbf{H}_k$ is predicted by an iteration-dependent convolutional neural network $N_{\boldsymbol{\theta}_{k}}$ with learnable parameters $\boldsymbol{\theta}_{k}$ that takes the current estimate $\mathbf{C}_k$ and the data fidelity gradient $\mathbf{g}_k$ as inputs, i.e., $\mathbf{H}_k = N_{\boldsymbol{\theta}_{k}}(\mathbf{C}_k,\mathbf{g}_k)$. The gradient $\mathbf{g}_k$ is computed following the FWI procedure using $\mathbf{M}_{\text{obs}}$ and $\mathbf{C}_k$ 
according to~\eqref{eq: get_grad}.
This formulation expresses the inversion as a sequence of learned, gradient-like updates.

The initial input $\mathbf{C}_0$ is defined as a homogeneous background map, representing a uniform water medium. To address the accumulated complexity of the inversion process, each iteration of the unfolding process uses a network $N_{\boldsymbol{\theta}_k}$ with the same architecture and different parameters. This design allows early iterations to make substantial updates, focusing on capturing broader features, while later iterations refine finer structures and make precise adjustments based on the gradient of the data fidelity at each step. 
Unlike MB-QRUS, which is essentially a single update and may oversmooth fine structures, DUFWI enables iterative learned correction, preserving the FWI mechanism while learning the update rule.
A visual illustration of the DUFWI method is shown in Fig.~\ref{fig: classical}(c).

\vspace{-0.43cm}
\subsection{Block-wise Training Process}
\vspace{-0.16cm}
The total training process begins by a training dataset $ \left\{ \bigl( \mathbf{M}_{\text{obs}}^{(i)}, \mathbf{C}_{gt}^{(i)} \bigr) \right\}_{i=1}^{M}$,
where each pair consists of observed CD and ground truth (GT) SoS maps,
%defines an inverse problem and its corresponding target solution for the unfolding algorithm, 
and \(M\) denotes the total number of training examples. In the $k$-th iteration, $\tilde{F}(\mathbf{C}_{k-1})$ is first obtained by performing wave propagation using previous estimate $\mathbf{C}_{k-1}$. The gradients $\mathbf{g}_k$ are precomputed offline for the entire training set by a single pass of the forward operator $\tilde{F}$ and back projection. $N_{\boldsymbol{\theta}_{k}}$ is trained in a supervised manner using the prepared gradient-update pairs.
In the \(k\)-th iteration, for each training sample $i$, the network-generated update $\mathbf{H}_k^{(i)} = N_{\boldsymbol{\theta}_{k}}(\mathbf{C}_k^{(i)},\mathbf{g}_k^{(i)})$ is added to $\mathbf{C}_k^{(i)}$ as in~\eqref{eq:dufwi_update}, thereby updating
current guess toward the $\mathbf{C}_{gt}^{(i)}$.
Accordingly, the training objective minimizes the discrepancy between $\mathbf{C}_{k+1}^{(i)}$ and the $\mathbf{C}_{gt}^{(i)}$ by minimizing the mean squared error (MSE) loss
\begin{equation} \label{eq: mse_loss}
	\mathcal{L}(\mathbf{\bm{\theta}}_k) = \frac{1}{M} \sum_{i=1}^M \Big\| \mathbf{C}_{gt}^{(i)} - \mathbf{C}_{k+1}^{(i)} \Big\|_2^2.
\end{equation}

In block-wise training, we train the current iteration neural network using the previous estimate and a single physics-based gradient computed with respect to the GT. After training, the network operates in inference mode to produce an updated estimate and computes the data-fidelity gradient from the physical model for each label. These outputs are used to prepare the training dataset for the subsequent iterative block. This avoids gradient backpropagation through the entire cascaded network, substantially reducing the computational overhead of training and enabling larger datasets.
%In addition, our approach reuses the precomputed GT and gradients during training, whereas end-to-end training necessitate computations at each iteration. This avoids repeated wave-propagation or adjoint computation, substantially reducing the computational overhead of training and making it feasible to train on larger datasets.
%In addition, our approach reuses the ground truth and corresponding CD during training, whereas end-to-end training necessitate simulating wave propagation at each iteration. 
%This design also ensures that the network keep accurately guided toward the global optimum in each iteration update. 
Compared to traditional FWI, although the gradient computation remains the same, this learning-based approach significantly accelerates convergence and has the potential to yield more accurate reconstructions.
% where \(\mathbf{\theta}_k\) represents the learnable parameters of the network \(N_{\mathbf{\theta}_k}\), and \(M\) is the size of the training dataset. At each iteration, the network is trained to produce an optimized update step that approximates the ground truth speed of sound map. This approach significantly accelerates convergence compared to traditional FWI and empirically results in more accurate final reconstructions.

\vspace*{-0.36cm}
\subsection{Deep neural network architecture}
\vspace{-0.1cm}
Following a similar architecture to~\cite{bn34}, the input is processed by three convolutional layers with an increasing number of output channels (64, 128, 256), each followed by a ReLU activation. The resulting feature maps are then concatenated and passed through another sequence of three convolutional layers with decreasing output channels (128, 64, 1). Only the first two layers in this second stage apply ReLU activations, while the final convolutional layer acts as the output layer of the network, producing the update $\mathbf{H}_k$. This output is added to $\mathbf{C}_k$ to generate the updated SoS map, as defined in \eqref{eq:dufwi_update}.
The convolutional layers employ $5 \times 5$ kernels, and zero-padding is applied to preserve the spatial dimensions of the input throughout the network. 
\begin{figure}[t!]
	\centering
	\includegraphics[width=0.49\textwidth, keepaspectratio]{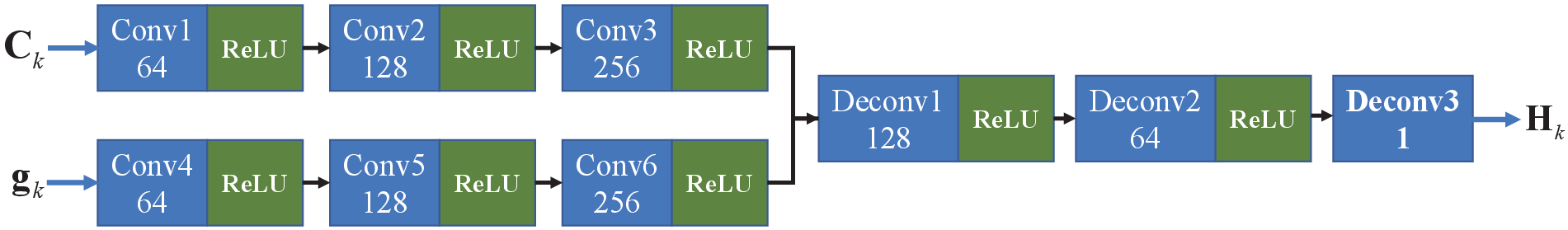}
	\vspace*{-1.36em}
	\caption{Illustration of the overall CNN architecture.}
	\label{fig: cnn}
	\vspace*{-1.66em}
\end{figure}

\vspace{-0.36cm}
\subsection{Deep neural network inference}
\vspace{-0.1cm}
After the training stage, the DUFWI framework can be applied during the inference within an iterative update process. 
In the $k$-th iteration, the gradient of the data misfit with respect to the SoS map is computed and fed into the corresponding trained neural network $N_{\boldsymbol{\theta}_k}$, which produces the update for that iteration. This process continues sequentially over all iterations.
DUFWI acts as a operator that maps the current gradient information and existing estimation to a more effective update step.  
The dominant cost during inference remains the gradient computation from the physical simulator, and the additional cost introduced by the DUFWI neural networks is negligible in comparison.
The networks are lightweight, resulting in minimal overhead both in terms of computational complexity and memory usage.

%\vspace{-0.39cm}
\section{Hardware System}\label{sec:harware system}
\vspace{-0.1cm}

We demonstrate our algorithm on hardware acquired CD measurements. To achieve this, we employed a Verasonics research US system (Verasonics, Inc., Kirkland, WA, USA) with 16 individually addressable custom immersion transducers (Blatek Industries, USA)
%300 kHz center frequency
%, 7.5 mm active diameter, dual matching layers)
arranged in a circular configuration with a 10 cm aperture diameter. The Verasonics platform allows the acquisition and storage of CD.

\begin{figure}[!h]
	\vspace*{-0.56em}
	\centering
	\includegraphics[width=0.156\textwidth, keepaspectratio]{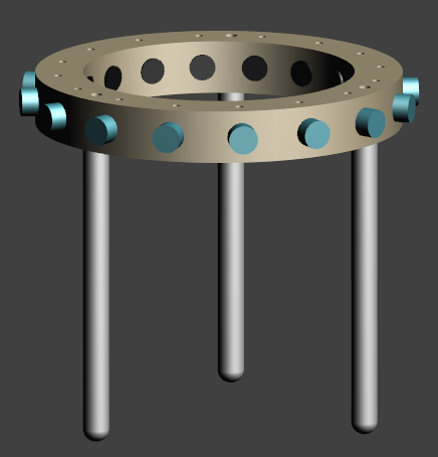}
	\vspace*{-0.3em}
	\caption{Layout of the transducer elements. The circular array consists of 16 evenly spaced elements arranged on a 10 cm diameter ring.}  
	\label{fig:transducer}
	\vspace*{-0.96em}
\end{figure}

\begin{figure}[!h]
	\vspace*{-0.6em}
	\centering
	\includegraphics[width=0.416\textwidth, keepaspectratio]{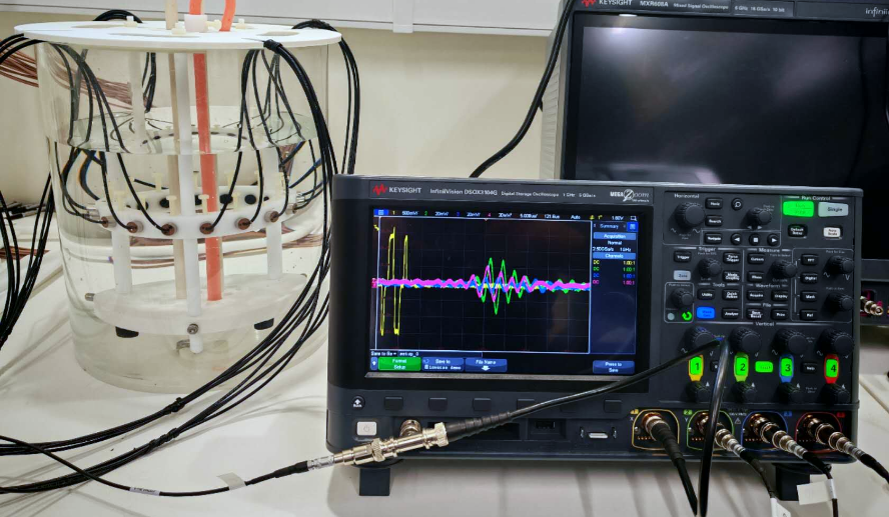}
	\vspace*{-0.3em}
	%	\caption{Overall Hardware System. On the left, a transparent water bath houses the circular transducer array. On the right, a Keysight InfiniiVision oscilloscope is used only to illustrate example received CD waveforms in real time.}
	\caption{Overall hardware system: (Left) a transparent water bath housing the circular transducer array; (Right) a Keysight InfiniiVision oscilloscope used only to illustrate example received CD waveforms in real time.}
	\label{fig:hardware}
	\vspace*{-0.86em}
\end{figure}

The arrangement of the transducer elements and the overall hardware system setup are illustrated in Fig.~\ref{fig:transducer} and Fig.~\ref{fig:hardware}, respectively. 
For each transmission, a single element was used as the transmitter, and three elements were used as receivers: the element positioned directly opposite the transmitter and its two adjacent elements.
This receive pattern yields $n_R = 3$ channels per transmission. Over $n_p=16$ transmissions, a total of $n_p \times n_R = 48$ channels are recorded.

\begin{figure}[h]
	\centering
%	\vspace*{-0.6em}
	\includegraphics[width=0.376\textwidth, keepaspectratio]{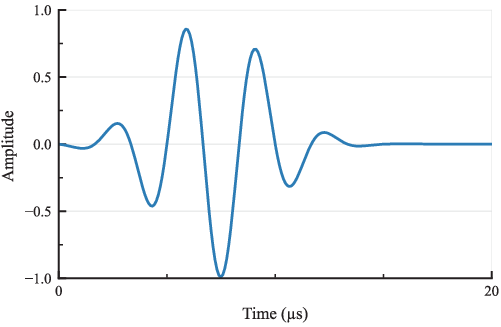}
	\vspace*{-0.8em}
%	\caption{The \hl{transmission} pulse waveform from Verasonics workspace.}
	\caption{Transmission pulse waveform from the Verasonics workspace. 
%		This figure shows only the first 20 $\mu$s of the waveform, the signal is zero thereafter.
	}
%		which is used for the hardware experiment}  
	\label{fig:pulse_shape}
	\vspace*{-1.96em}
\end{figure}

The transmission source pulse for each element is a Gaussian pulse with an center frequency of nominally 350 kHz, determined by the transducer hardware.
The pulse shape is shown in Fig.~\ref{fig:pulse_shape}. 
For hardware acquisition, the receive waveform is sampled at \(f_s = 62.5\,\text{MHz}\) over a \(112\,\mu\text{s}\) capture window, yielding 7000
%\(f_s \times 112\,\mu\text{s} = 7000\) 
uniformly spaced samples per channel. This window exceeds the simulated pulse duration and provides margin for subsequent alignment and spectral calibration.

Array calibration is essential for high-performance US imaging because it corrects systematic discrepancies between the acoustic fields produced by individual transducer elements and those predicted by physics-based numerical models. To align simulated and measured CD, we calibrate the transmitted source pulse following~\cite{Calib}. Let $s_{\rm sim}(t)$ denote the simulated source pulse, $r^{(i)}_{\rm sim}(t)$ the simulated receive CD on the $i$-th Tx-Rx path, and $r_{\rm hw}^{(i)}(t)$ the corresponding measured CD. Assuming a linear time-invariant (LTI) propagation/receive chain with frequency response $H^{(i)}(f)$, we have $R_{\rm sim}^{(i)}(f) = H^{(i)}(f)S_{\rm sim}(f)$ and $R_{\rm hw}^{(i)}(f) = H^{(i)}(f)S^{(i)}_{\rm hw}(f)$, where $S_{\rm sim}(f)$, $R_{\rm sim}^{(i)}(f)$, and $R_{\rm hw}^{(i)}(f)$ denote the Fourier transforms with respect to the ordinary frequency $f$ (in Hz) of $s_{\rm sim}(t)$, $r_{\rm sim}^{(i)}(t)$, and $r_{\rm hw}^{(i)}(t)$, respectively. Eliminating $H^{(i)}(f)$ yields the calibration mapping $S_{\rm hw}^{(i)}(f) = S_{\rm sim}(f)\frac{R_{\rm hw}^{(i)}(f)}{R_{\rm sim}^{(i)}(f)}$. Consequently, the calibrated time-domain pulse is obtained by applying the inverse Fourier transform to the calibrated spectrum, i.e., $s_{\rm hw}^{(i)}(t) = \mathcal{F}^{-1}\bigl\{S_{\rm hw}^{(i)}(f)\bigl\}$.

In practice, $s_{\rm sim}(t)$ is chosen as the time derivative of a Gaussian pulse and is discretized with \(N_s = 1034\) time samples at a sampling interval \(\Delta t_s = 90.243\,\text{ns}\) (total duration \(N_s \Delta t_s \approx 93.31\,\mu\text{s}\)). The simulated receive signals $r^{(i)}_{\rm sim}(t)$ are generated by numerically solving US wave propagation on a 10 cm × 10 cm computational domain matched to the transducer array dimensions. 
The region is discretized on a uniform grid with spacing $\lambda / 8$ (where $\lambda$ is the wavelength), resulting in a $188 \times 188$ grid. 
%The Courant–Friedrichs–Lewy (CFL) condition is verified to ensure numerical stability and convergence to a valid PDE solution~\cite{bn35}. 
Numerical stability and convergence to a valid PDE solution are ensured by verifying the Courant–Friedrichs–Lewy (CFL) condition.
%{\color{blue}A PML of 20 grid points is applied on each boundary, which is kept fixed across all experiments. We use a quadratic damping profile within the PML with a target attenuation factor $10^{-3}$, yielding a maximum damping coefficient of $1.46 \times 10^6$ at the outer boundary.}
A PML of 20 grid points is applied on each boundary. We use a quadratic damping profile within the PML with a target attenuation factor $10^{-3}$, yielding a maximum damping coefficient of $1.46 \times 10^6$ at the outer boundary. Both the PML thickness and damping parameters are fixed across all experiments.

To improve numerical conditioning and temporal localization, we apply the calibration to the differentiated source wavelet. Specifically, Let \(s_{\rm diff}(t)=\frac{d}{dt}s_{\rm sim}(t)\), which corresponds to \(S_{\rm diff}(f)=j2\pi f\,S_{\rm sim}(f)\) in the frequency domain, where $j = \sqrt{-1}$. We then estimate the equivalent hardware source via \(S_{\rm hw}(f)=S_{\rm diff}(f)\,R_{\rm hw}^{(i)}(f)/R_{\rm sim}^{(i)}(f)\) and \(s_{\rm hw}^{(i)}(t)=\mathcal{F}^{-1}\{S_{\rm hw}(f)\}\). The resulting wavelet is zero-mean with a well-localized main lobe and reduced long-duration ringing. Note that differentiation only changes the overall amplitude scaling,
%and does not bias calibration, 
since any global scaling of the simulated source also appears in $R^{(i)}_{\rm sim}(f)$ and cancels in spectral ratio.

\begin{figure}[t]
	\vspace*{-0.5em}
	\centering
	\subfigure[]{%
		\includegraphics[width=0.241\textwidth]{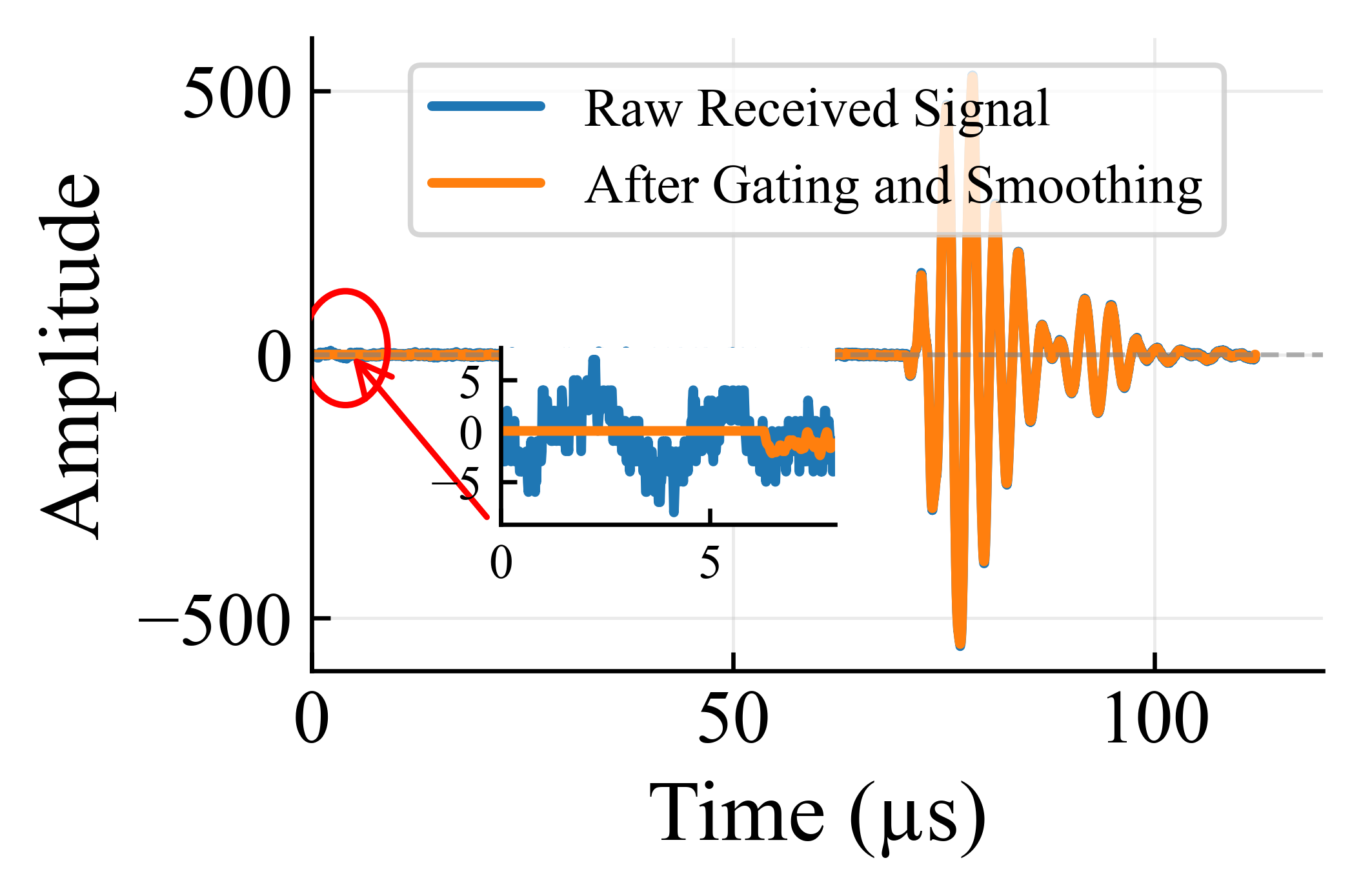}%
		\label{figcb1}}
	\hfill
	\subfigure[]{%
		\includegraphics[width=0.241\textwidth]{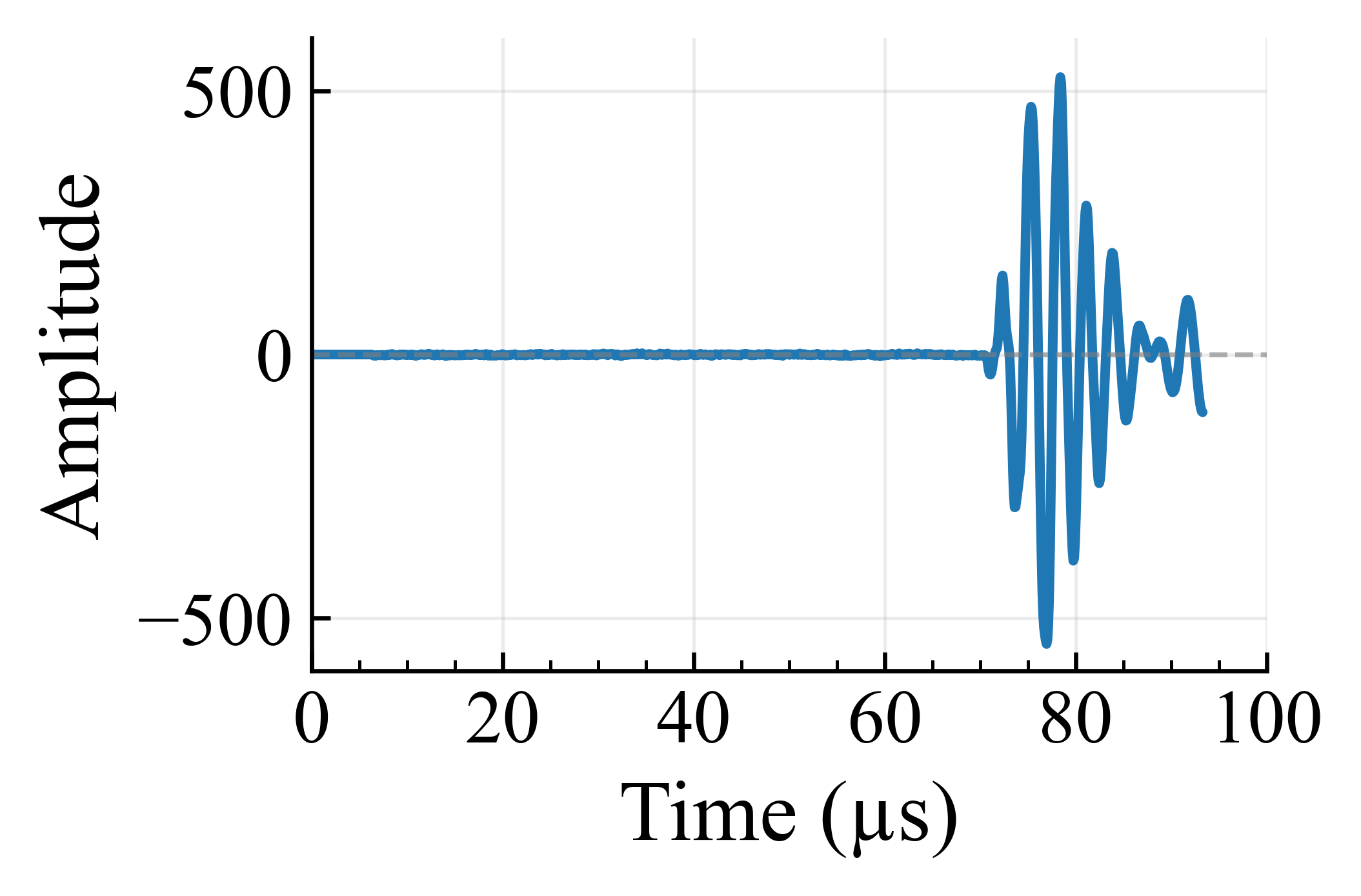}%
		\label{figcb2}}
	\\ \vspace*{-0.5em}
	\subfigure[]{%
		\includegraphics[width=0.241\textwidth]{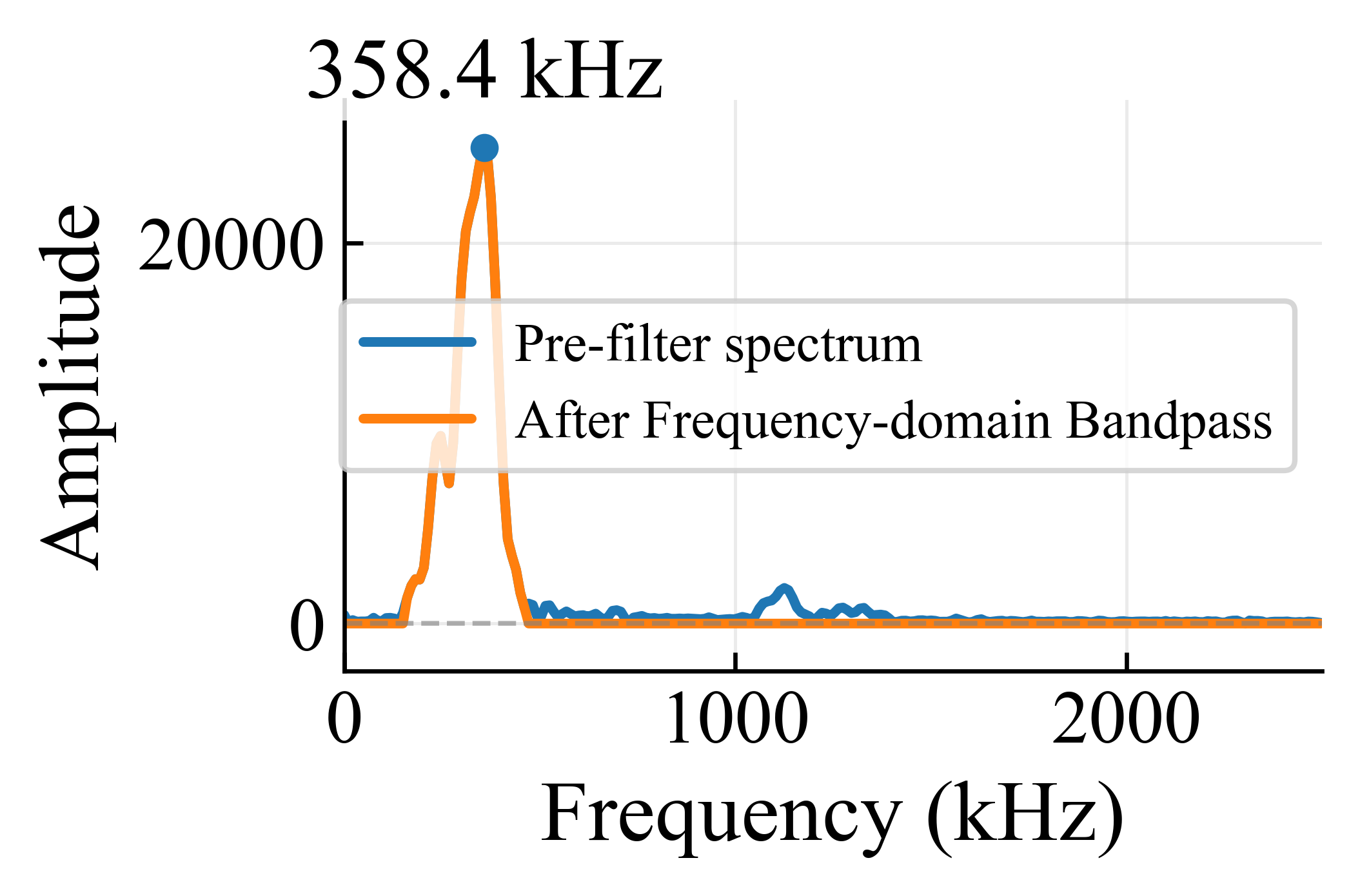}%
		\label{figcb3}}
	\hfill
	\subfigure[]{%
		\includegraphics[width=0.241\textwidth]{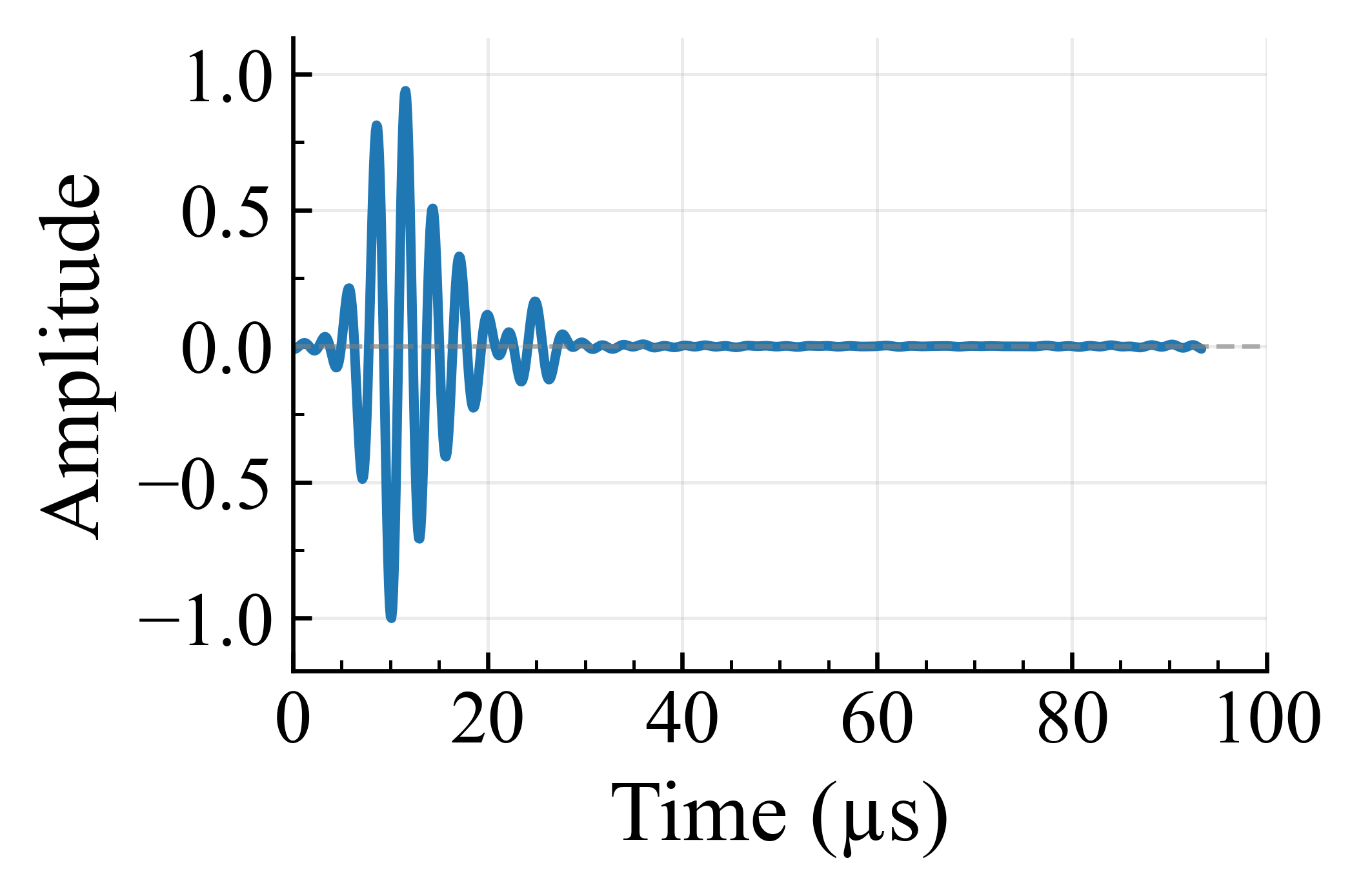}%
		\label{figcb4}}
	\caption{Illustration of the calibration process. (a) raw vs. gated and smoothed recorded signal; (b) the signal interpolated to the simulation time grid; (c) magnitude spectra before and after the frequency-domain bandpass; (d) normalized calibrated source pulses.}
	\label{fig:cb}
	\vspace*{-1.8em}
\end{figure}

At the beginning of the received signals, strong reflections arise from the probe lenses and the phantom interface.  
Therefore the initial $400$ time samples are set to zero to remove such interference,
and a moving average filter with a window size of 10 samples is applied, followed by temporal alignment through a shift of 5 samples. A comparison of the raw and processed measured singal is shown in Fig.~\ref{fig:cb}(a).
Subsequently, the filtered signals are digitally resampled (via interpolation) to match the temporal sampling rate of the simulations, as shown in Fig.~\ref{fig:cb}(b). A frequency-domain bandpass filter is then applied to these resampled signals to preserve the main spectral components of the source pulse and attenuate out-of-band noise, as shown in Fig.~\ref{fig:cb}(c).

Finally, the ratio between the measured signal spectrum and the simulated received signal spectrum is computed and multiplied by the simulated source pulse spectrum, which produces an estimate of the hardware source pulse. The calibrated source pulse is recovered by applying the inverse Fourier transform and then peak-normalized to unit magnitude for consistency across channels and to prevent amplitude saturation. 
The calibrated source are shown in Fig.~\ref{fig:cb}(d). This calibration aligns the simulated and measured CD in both timing and spectrum, which is critical for stable inversion.

\vspace{-0.39cm}
\section{Numerical Results} \label{subsec: results} 
\vspace{-0.1cm}

The inversion methods described in Section~\ref{sec: problem_statement} are first validated on simulated datasets, followed by evaluation on measured hardware data. 
Specifically, two types of simulated datasets are employed for subsequent evaluations, with the two networks are trained independently.

\textit{1) MNIST dataset:}
First, the SoS maps dataset used for training is derived from the MNIST~\cite{MNIST}.
%Reconstructing MNIST digits can be challenging because the cavities (hollow or enclosed regions) within the digits act as resonators, which leads to strong multiple scattering phenomena that increase the nonlinearity of the inverse problem.
%The reconstruction of MNIST digits is challenging, as the hollow or enclosed regions inside the digits may serve as resonators, resulting in strong multiple scattering and increased nonlinearity of the inverse problem.
The reconstruction of MNIST digits is challenging when enclosed regions are present, because such regions may exhibit resonance effects that strengthen multiple scattering and increase the nonlinearity of the inverse problem.
To prepare this dataset, we load each digit image and apply a sequence of transformations: conversion to tensor (normalizing pixel values to $[0,1]$), random rotations ($\pm30^\circ$), translations ($\pm5$ pixels), scaling ($0.8\times$–$1.2\times$), shear ($\pm10^\circ$), and random erasing. Each augmented image is then multiplied by a random factor sampled uniformly from $[0.3,\,1.2]$, thereby exposing the network to a broad range of SoS contrasts. Next, the image is resized to fit the circular transducer array grid. 
A final random in-plane rotation is applied before scaling the image intensities by 666\,m/s and adding a bias of 1550\,m/s to yield the GT.
% SoS map. 

\textit{2) Simulated arm dataset:}
To evaluate the performance of the proposed method on arm scenario, we generated a simulated arm dataset designed to mimic the cross-sectional profile of a human arm. Each sample is defined on a $10~\mathrm{cm}\times10~\mathrm{cm}$ rectangular domain and contains an elliptical foreground region in which tissue properties are assigned. The outer boundary represents skin. 
Beneath the skin lies subcutaneous fat.
The interior is muscle compartment. Two circular bone regions are placed in the muscle. An optional small elliptical edema region appears in the muscle. To train and evaluate the network, we constructed a single combined dataset that includes both edema-present and edema-absent samples, and the dataset was balanced with a $1\!:\!1$ ratio between the two groups, and the network was trained on this combined set rather than on two separate datasets. Across samples, arms differ in their external contour and their underlying tissue structures, including differences in skin, bone, and edema.
The SoS values for each tissue type were selected based on physiological measurements, including muscle (1570--1620 m/s), fat (1420--1450 m/s), skin (1530--1560 m/s), bone (2700--3000 m/s), and edema (1450--1500 m/s)~\cite{arm_data}.

In training process, the proposed DUFWI framework is unrolled for \(5\) iterations. At each iteration \(k\), we train the corresponding network \(N_{\bm{\theta}_k}\) for 40 epochs using a batch size of 32 and the Adam optimizer~\cite{adam}. The learning rate for different network in each iteration is set according to the schedule $ \eta_k = [1e^{-4},\;1e^{-4},\;5e^{-5},\;1e^{-5},\;1e^{-5}]$. The training loss for DUFWI and MB-QRUS is the MSE. All other training hyperparameters remain fixed across iterations. 
In addition, we apply an offset-and-normalization to the input SoS map for numerical stability. We subtract 1400 m/s as a soft-tissue reference near the lower end of dataset. Although bone has much larger SoS, the subsequent normalization maps the dynamic range to comparable magnitudes for the network. After the network output, a corresponding post-processing step rescales the output $\mathbf{H}_{k}$ to keep align with the original units.

All models (DUFWI and MB-QRUS) were trained using one NVIDIA A40 GPU equipped with 70 GB memory, while the experiments were conducted in PyTorch 2.5.1. 
%Training of all models (DUFWI and MB-QRUS) was carried out using one NVIDIA A40 GPU equipped with 70 GB memory, while the experiments were developed in PyTorch 2.5.1.
To assess MB-QRUS on our problem setup, we retrained the network from Section~\ref{subsec: MB-QRUS} using our SoS maps and forward model, and adapted it to our 16‑element circular array configuration. 
In addition, for the FWI baseline across all experiments, we implement an ADMM-based solver with 200 outer iterations (no stopping criterion) and an $\ell_1$-type regularization applied to the spatial derivatives (total variation regularization). In each iteration, the data-fidelity subproblem is minimized using L-BFGS with learning rate is 1 and iteration is 5~\cite{AFWI1,AFWI2}.

For each type of simulated dataset, we generate 20,000 SoS maps, partitioned into 16,000 for training and 4,000 for validation. The SoS maps above described are used as the GT. 
To generate a training dataset $\left\{\left( \mathbf{M}_{\text{obs}}^{(i)}, \mathbf{C}_{gt}^{(i)}\right)\right\}$, the observed CD for each example is calculated using the forward model described in Section \ref{subsec:Problem Formulation} as \(\mathbf{M}_{\text{obs}}^{(i)} = \tilde{F}(\mathbf{C}_{gt}^{(i)})\), which is required by both DUFWI and MB\mbox{-}QRUS. All training datasets are noise-free, with additive gussian white noise being added to the CD during testing under signal-to-noise ratio (SNR) is 30 dB. This setup allows the model to learn the underlying properties from noise-free data during training, while evaluating the robustness to noise during testing.

\vspace{-0.36cm}
\subsection{MNIST Dataset Results}
\vspace{-0.1cm}
\begin{figure}[!t]
	\vspace*{-0.66em}
	\centering
	\includegraphics[width=0.49\textwidth]{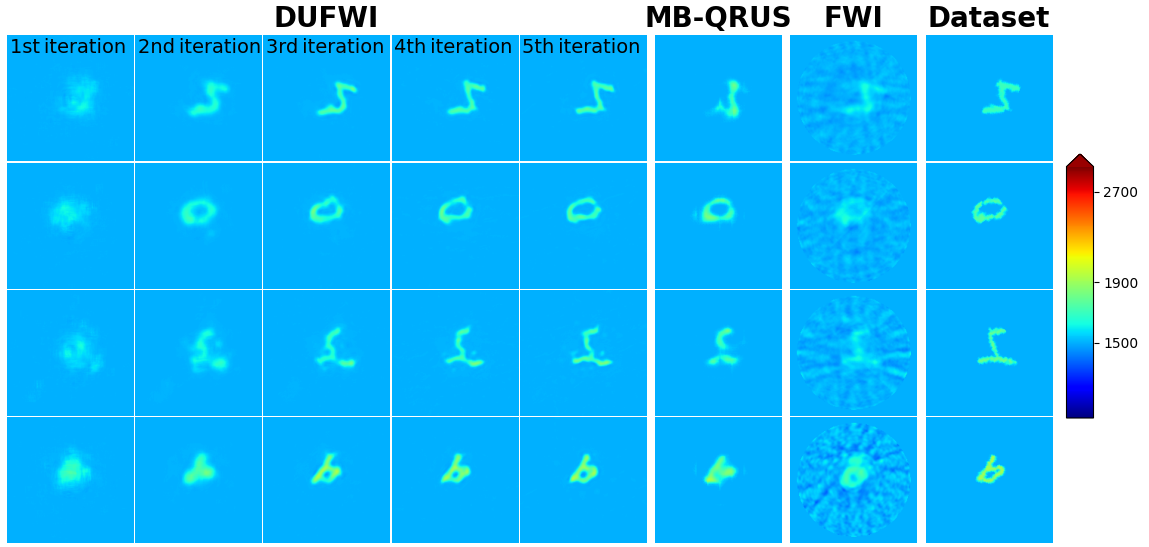}
%	\vspace*{-1.36em}
	\caption{Reconstructed SoS maps from the MNIST test subset, displayed across several unfolded iterations. This visualization shows the progression and refinement of the SoS maps through multiple iterations.}  
	\label{fig:Compare_all_minst}
	\vspace*{-1.66em}
\end{figure}

% We begin by presenting the results obtained from the MINST datasets. 
Fig.~\ref{fig:Compare_all_minst} illustrates the reconstructed SoS maps of DUFWI (shown after postprocessing) compared to MB-QRUS, FWI, and GT on four MNIST-based test examples. Across all methods, the initial SoS map $\mathbf{C}_0$ corresponds to a homogeneous water medium. DUFWI reconstructs finer structural details and yields visually more accurate SoS maps than MB\mbox{-}QRUS and conventional FWI.
%We find that our approach performs better than MB-QRUS and FWI on the SoS maps dataset we generated, specifically, it is able to reconstruct finer details. 
In addition, as illustrated in Fig.~\ref{fig:Compare_all_minst}, it can be observed that DUFWI captures the approximate location of object in the first iteration, constructs its overall shape in the second iteration, and then progressively refines the detailed image features in subsequent iterations. To better quantify the advantages of DUFWI, three metrics are used to evaluate the test‐set results: the structural similarity index (SSIM), the peak signal-to-noise ratio (PSNR), and the normalized mean squared error (NMSE) to gauge the accuracy of the reconstructed physical parameters\cite{bn8, metrics}. A full quantitative comparison across iterations appears in Table \ref{tab: minst_set}.
%, the proposed DUFWI method consistently outperforms MB-QRUS in quantitative metrics in the final iteration. 
Across all iterations, the proposed DUFWI method consistently outperforms FWI, and in the final iteration it also surpasses MB-QRUS on the quantitative metrics.
During the first two iterations, DUFWI exhibits relatively lower metric values, however, in the third iteration, its performance surpasses that of MB-QRUS. In subsequent iterations, DUFWI's metrics continue to improve, consistent with the behavior illustrated in Fig.~\ref{fig:Compare_all_minst}.
%In subsequent iteration, the metrics for DUFWI continue to improve.
%% but albeit at a slower rate. 
%This trend is consistent with the behavior illustrated in Fig.~\ref{fig:Compare_all_minst}.
% It is important to note that our method required only 5 gradient computations for reconstruction, whereas FWI needed 200, and MB-QRUS required one forward computation. 
\begin{table}[htbp]
	\vspace*{-1.36em}
	\centering
	\caption{Quantitative comparison on MNIST test data. The quantitative metrics of DUFWI improve with each iteration.
%		, demonstrating performance improvement over iterations. 
	Arrows indicate the preferred direction (\(\uparrow\) higher is better, \(\downarrow\) lower is better).}
	\vspace*{-0.5em}
	\label{tab: minst_set}
	\resizebox{\columnwidth}{!}{
		\begin{tabular}{|l|c|c|c|c|c|c|c|}
			\hline
			\multicolumn{8}{|c|}{\textbf{MNIST Test Data}} \\
			\hline
			\multirow{2}{*}{\textbf{Metric}} &
			\multirow{2}{*}{\textbf{\makecell{FWI}}} & \multirow{2}{*}{\textbf{\makecell{MB-\\QRUS}}} &
			 \multicolumn{5}{c|}{\textbf{DUFWI}} \\
			\cline{4-8}
			& & & \textbf{1st iter} & \textbf{2nd iter} & \textbf{3rd iter} & \textbf{4th iter} & \textbf{5th iter} \\
			\hline
			SSIM~(\(\uparrow\))     & 0.9409 & 0.9871 & 0.9787 & 0.9833 & 0.9877 & 0.9901 & \textbf{0.9916} \\
			PSNR (dB)~(\(\uparrow\))& 38.39 &  43.41 & 40.58  & 43.22  & 44.59  & 46.58  & \textbf{47.41} \\
			NMSE (dB)~(\(\downarrow\)) &  -33.98 & -36.47 & -34.13 & -35.63 & -37.00 & -38.01 & \textbf{-38.81} \\
			\hline
		\end{tabular}%
	}
\vspace*{-1.36em}
\end{table}

\vspace{-0.36cm}
\subsection{Simulated Arm Dataset Results}
\vspace{-0.1cm}
\begin{figure}[t!]
	\vspace*{-0.8em}
	\centering
	% (a) without edema
	\subfigure[Simulated arm without edema\label{fig:Compare_all_no_edema}]{
		\includegraphics[width=0.439\textwidth,keepaspectratio]{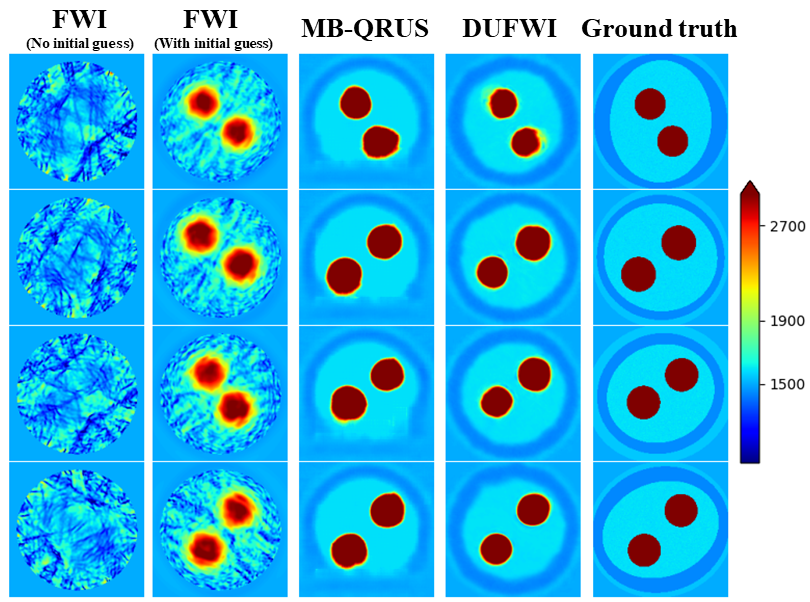}
	}
	
	% (b) with edema
	\subfigure[Simulated arm with edema\label{fig:Compare_all_edema}]{
		\includegraphics[width=0.443\textwidth,keepaspectratio]{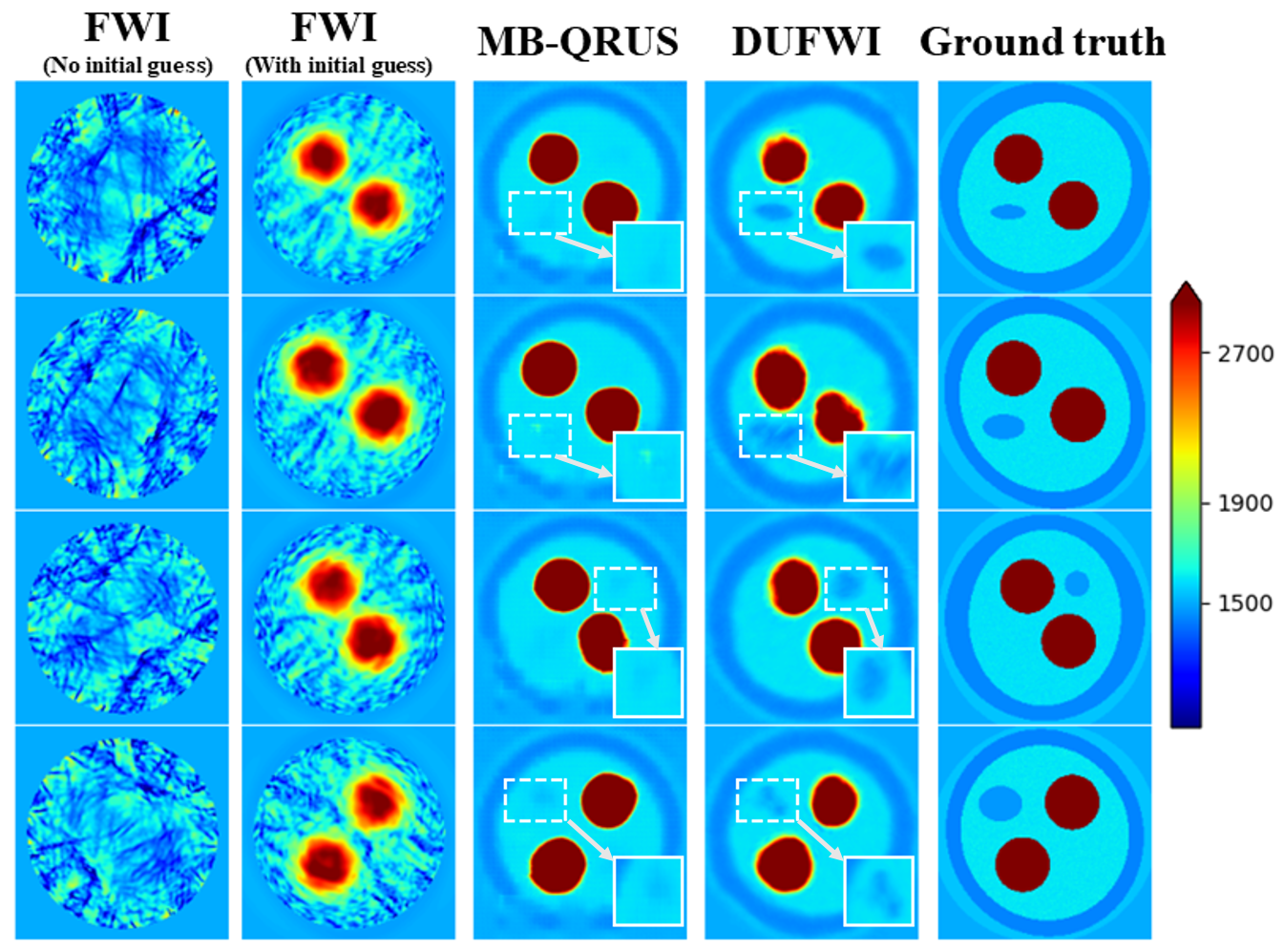}
	}
	\vspace*{-0.39em}
	\caption{Reconstructed SoS maps of simulated arm examples: (a) non-edema case and (b) edema case. The figure compares the performance of DUFWI with FWI and MB-QRUS. 
%		This visualization also shows the edema classification results by DUFWI and MB-QRUS.
	}
	\label{fig:Compare_all}
	\vspace*{-2.06em}
\end{figure}

%	To evaluate the effectiveness of the proposed DUFWI method in anatomical applications, two complementary simulated arm datasets were generated: a non-edema case consisting of two bones embedded within a uniform soft-tissue background, and an edema case incorporating an elliptical, low-speed region to simulate the accumulation of interstitial fluid accumulation. 
To further test the effectiveness of the DUFWI in anatomical applications, we evaluate all methods on simulated arm test data with and without edema. Owing to the stronger tissue heterogeneity and SoS contrast in this dataset, conventional FWI stagnate or fail under unfavorable initial conditions. Consequently, we benchmark the proposed DUFWI against MB-QRUS and conventional FWI under two complementary initialization strategies: one employs an initial SoS $\mathbf{C}_0$ assuming a homogeneous water medium consistent with DUFWI and MB-QRUS, while the alternative initial guess incorporates approximate bone locations embedded within a water background through Gaussian blurring, leaving the SoS and positions of muscle, skin, and edema unspecified. 

%Further, the proposed DUFWI method is compared with the MB-QRUS method and conventional FWI implemented within an ADMM framework under two different initialization strategies: one strategy employs an initial SoS $\mathbf{C}_0$ assuming a homogeneous water medium consistent with DUFWI and MB-QRUS, while the alternative initial guess incorporates approximate bone locations embedded within a water background through Gaussian blurring, leaving the SoS and positions of muscle, skin, and edema unspecified. 

Fig.~\ref{fig:Compare_all} summarizes the reconstructed SoS maps for both scenarios: the non-edema in Fig.~\ref{fig:Compare_all_no_edema} and the edema in Fig.~\ref{fig:Compare_all_edema}. 
%Fig.~\ref{fig:Compare_all} illustrates the reconstructed SoS maps of simulated arm test data for the non-edema case in Fig.~\ref{fig:Compare_all_no_edema} and edema case in Fig.~\ref{fig:Compare_all_edema}. 
When initialized from a homogeneous water medium, conventional FWI produces severe speckle artifacts, fails to recover the high-speed bone tissue and unable to reconstruct the low‐speed skin boundaries or the internal edema region in the edema case. Although the introduction of an initial guess significantly reduces background artifacts and sharpens the bone interfaces, skin boundaries remain blurred and the edema region remains undetectable. The MB-QRUS method yields nearly speckle-free and smooth reconstructions, yet this  oversmoothing distorts skin morphology and nearly eliminates visibility of the low-speed edema region, indicating that MB-QRUS may converge to a local minimum. 
\begin{table}[!t]
	\vspace*{-1.0em}
	\caption{Quantitative Comparison for the Reconstructed SoS Maps of Simulated Arm Data for Non-Edema Case Shown in Fig.~\ref{fig:Compare_all_no_edema}.}
	\vspace*{-0.5em}
	\centering
	\resizebox{\columnwidth}{!}{
		\begin{tabular}{|l|c|c|c|c|}
			\hline
			\multicolumn{5}{|c|}{\textbf{Simulated Arm Data for Non-Edema Case}}\\
			\hline
			\multirow{2}{*}{\textbf{Metric}} & \multicolumn{2}{c|}{\textbf{FWI}} & \multirow{2}{*}{\textbf{MB-QRUS}} & \multirow{2}{*}{\textbf{DUFWI}} \\
			\cline{2-3}
			& No Init. Guess & With Init. Guess &  & \\
			\hline 
			SSIM~(\(\uparrow\)) & 0.5894 & 0.7187 & 0.9138 & \textbf{0.9163} \\
			%			\hline
			PSNR (dB)~(\(\uparrow\)) & 15.4119 & 20.8215 & 29.0541 & \textbf{29.4148} \\
			%			\hline
			NMSE (dB)~(\(\downarrow\)) & -12.2213 & -16.8712 & -23.1136 & \textbf{-23.3365} \\
			\hline
		\end{tabular}
	}
	\label{tab:quantitative_non_edema}
	\vspace*{-1.2em}
\end{table}

\begin{table}[!t]
%	\vspace*{-0.3em}
	\caption{Quantitative Comparison for the Reconstructed SoS Maps of Simulated Arm Data for Edema Case Shown in Fig.~\ref{fig:Compare_all_edema}.}
		\vspace*{-0.5em}
	\centering
	\resizebox{\columnwidth}{!}{
		\begin{tabular}{|l|c|c|c|c|}
			\hline
			\multicolumn{5}{|c|}{\textbf{Simulated Arm Data for Edema Case}}\\
			\hline
			\multirow{2}{*}{\textbf{Metric}} & \multicolumn{2}{c|}{\textbf{FWI}} & \multirow{2}{*}{\textbf{MB-QRUS}} & \multirow{2}{*}{\textbf{DUFWI}} \\
			\cline{2-3}
			& No Init. Guess & With Init. Guess &  & \\
			\hline 
			SSIM~(\(\uparrow\)) & 0.5853 & 0.7095 & 0.8957 & \textbf{0.9059} \\
			%			\hline
			PSNR (dB)~(\(\uparrow\)) & 15.5052 & 20.2911 & 28.8906 & \textbf{29.1282} \\
			%			\hline
			NMSE (dB)~(\(\downarrow\)) & -12.1594 & -16.7023 & \textbf{-23.9019} & -23.8702 \\
			LMSE ~(\(\downarrow\)) & 526.1629 & 361.7628 & 380.9106 & \textbf{255.1932} \\
			\hline
		\end{tabular}
	}
	\label{tab:quantitative_edema}
		\vspace*{-2.3em}
\end{table}
\begin{figure}[h]
	\vspace*{-0.5em}
	\centering
	\includegraphics[width=0.356\textwidth]{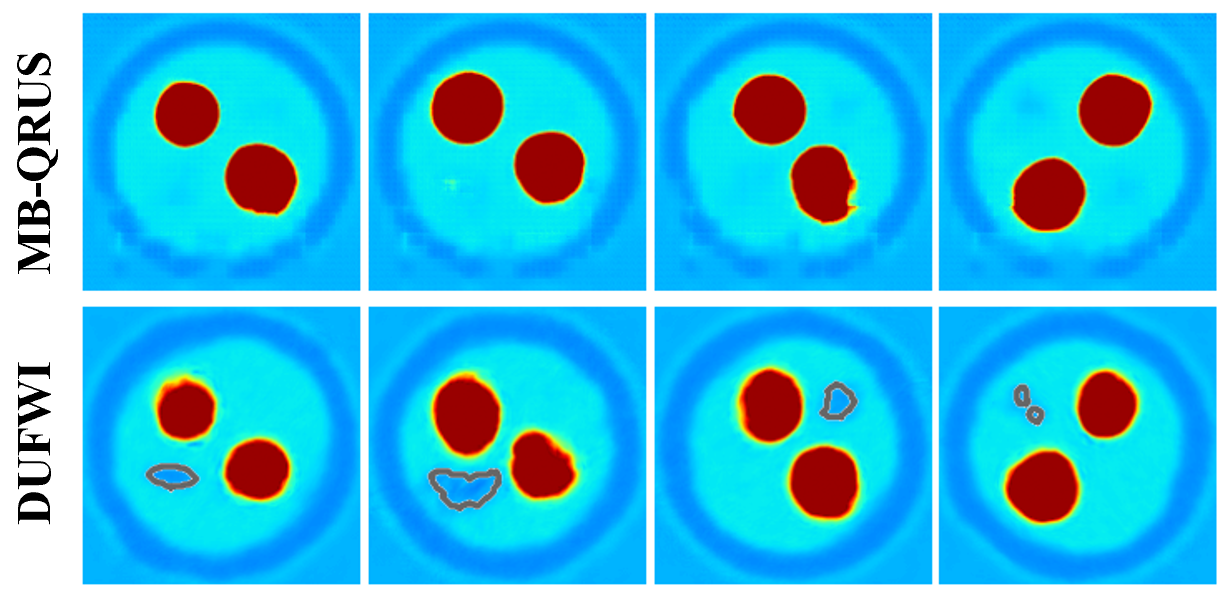}
	\vspace*{-0.36em}
	\caption{Classification results for the Simulated Arm Data in Fig.~\ref{fig:Compare_all_edema} obtained by DUFWI (last iteration) and MB-QRUS.
	}
	\label{fig:class_res}
	\vspace*{-0.5em}
\end{figure}

In contrast, DUFWI employs multiple unfolding iterations that enforce data fidelity together. In the non-edema case, DUFWI clearly resolves bone regions, while in the edema case, it accurately delineates both high-speed bones and surrounding low-speed tissues.
%, maintaining contrast and geometry closely matching the GT. 
Compared with MB-QRUS, DUFWI better reconstructs low-speed tissues adjacent to bone in the edema scenario, highlighting its potential for medical imaging applications aimed at diagnosing and characterizing abnormal tissues. The quantitative comparison of non-edema case and edema case are provided in Table~\ref{tab:quantitative_non_edema} and Table~\ref{tab:quantitative_edema}, respectively. In the non-edema case, the proposed DUFWI method outperforms MB-QRUS across all quantitative metrics, demonstrating superior reconstruction accuracy. 
Additionally, in the edema case, MB-QRUS shows marginally lower NMSE than DUFWI, whereas DUFWI maintains higher SSIM and PSNR. As a global squared-error metric, NMSE is largely driven by the majority non-edema region.
%Additionally, in the edema case, MB-QRUS achieves a slightly better NMSE compared to DUFWI, whereas the proposed DUFWI keeps higher SSIM and PSNR. This can be attributed to the mechanism of the NMSE. NMSE quantifies a global squared amplitude discrepancy and is dominated by the large population of non-edema pixels.
Although MB-QRUS produces smoother reconstructions, it tends to oversmooth low-speed regions like edema. We further report local MSE (LMSE) to evaluate the reconstruction accuracy inside the edema region, where DUFWI achieves the lowest LMSE, indicating a more accurate recovery of edema.

Besides, since both DUFWI and MB\text{-}QRUS recover high-contrast structures (e.g., bone and skin) comparably well, we further probe the low-contrast edema region in Fig.~\ref{fig:class_res}. To diagnose the edema, we introduce a classification procedure for edema detection on reconstructed SoS maps. After excluding bone and skin masks, we apply Gaussian smoothing to the remaining tissue region to suppress high-frequency noise while preserving low-contrast structures. Candidate edema regions are obtained by thresholding the SoS map to the physiologically informed range of 1460–1500, and then 2-D connected-component labeling on the binarized mask is performed, discarding components that fail basic plausibility checks (e.g., minimum area or proximity to excluded bone/skin masks). Finally, for each component we extract a sub-pixel contour and use the resulting closed curves as the edema boundaries. This classification method is performed on both DUFWI and MB\text{-}QRUS.
Applying this procedure to all test samples, we obtain edema classification quantitative results summarized in Table~\ref{tab:quantitative_edema_classification}, which show that DUFWI achieves markedly better diagnostic performance for edema classification than MB\text{-}QRUS. Specifically, DUFWI achieves 100\% precision (no false positives), while maintaining 88\% recall, indicating reliable detection of edema cases. In contrast, MB-QRUS attains 88.235\% precision but only 0.75\% recall, indicating a high false negative rate. This is mainly due to oversmoothing of MB-QRUS, which suppresses the low-speed contrast of the edema region. Consequently, its overall accuracy is substantially lower To complement the sample-level classification results, we further report the Dice coefficient in Table~\ref{tab:quantitative_edema_classification} to quantify the spatial overlap between the detected edema region and the reference edema mask derived from the GT. 
MB\text{-}QRUS attains a near-zero Dice, indicating negligible spatial overlap with the reference edema region. 
DUFWI improves the Dice to 0.5302, indicating improved edema localization and boundary delineation. As no visible edema can be observed in the FWI results (Fig.~\ref{fig:Compare_all_edema}), FWI appears ill-suited for edema detection, since it fails to provide discriminative edema-related features. Therefore, we omit the comparison with FWI in Table~\ref{tab:quantitative_edema_classification}.
Overall, DUFWI yields a more reliable edema assessment than MB\text{-}QRUS, supported by improved sensitivity and spatial agreement with the GT.

\begin{table}[!t]
	\vspace*{-0.86em}
	\caption{Edema Detection Performance on the Simulated Arm Dataset (Acc./Prec./Rec for both cases; Dice only for edema cases)}
	\vspace*{-0.5em}
	\centering
	{\fontsize{8pt}{9pt}\selectfont
	\begin{tabular}{|l|c|c|}
		\hline
		\multicolumn{3}{|c|}{\textbf{Simulated Arm Dataset for Both Case}}\\
		\hline
		\textbf{Metric} & \textbf{MB-QRUS} & \textbf{DUFWI} \\
		\hline
		Accuracy~(\%) &  50.325  &  \textbf{94.0}  \\
		Precision~(\%) &  88.235  &  \textbf{100}  \\
		Recall~(\%) &  0.75  &  \textbf{88.0}  \\
		Dice~(\(\uparrow\)) &  0.0355  &  \textbf{0.5302}  \\
		\hline
	\end{tabular}
	\label{tab:quantitative_edema_classification}
	\vspace*{-0.26em}
	}
\end{table}
\begin{figure}[!t]
	\vspace*{-0.33em}
	\centering
	\includegraphics[width=0.396\textwidth]{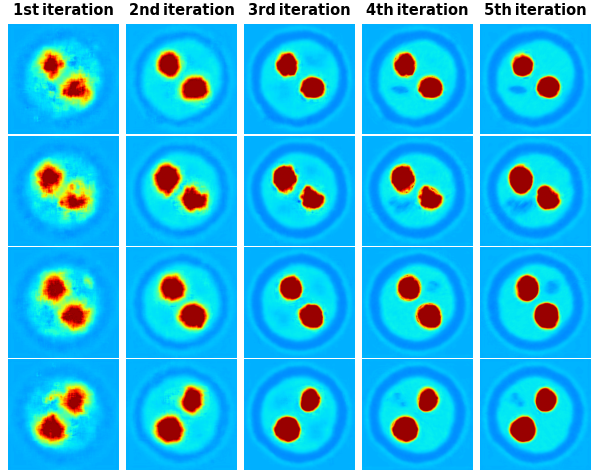}
	\vspace*{-0.3em}
	\caption{Reconstructed SoS maps from the simulated arm with edema examples, displayed across 5 unfolded iterations. }
	\label{fig:iter_dufwi}
	\vspace*{-1.86em}
\end{figure}

Fig.~\ref{fig:iter_dufwi} illustrates the iterative evolution of DUFWI for SoS reconstruction in edema case. During the first two iterations, DUFWI primarily recovers high-contrast anatomical boundaries. Subsequently, DUFWI progressively recovers the low-contrast edema region. This progression indicates that, compared with MB-QRUS, DUFWI is less susceptible to suboptimal local minima and yields more faithful SoS estimates within and around the edema, beneficial for medical diagnosis.

\begin{figure*}[!t]
%	\vspace*{-1.2em}
	\centering
	% (a) without edema
	\subfigure[\label{fig:res_hardware_1}]{
		\includegraphics[width=0.75\textwidth,keepaspectratio]{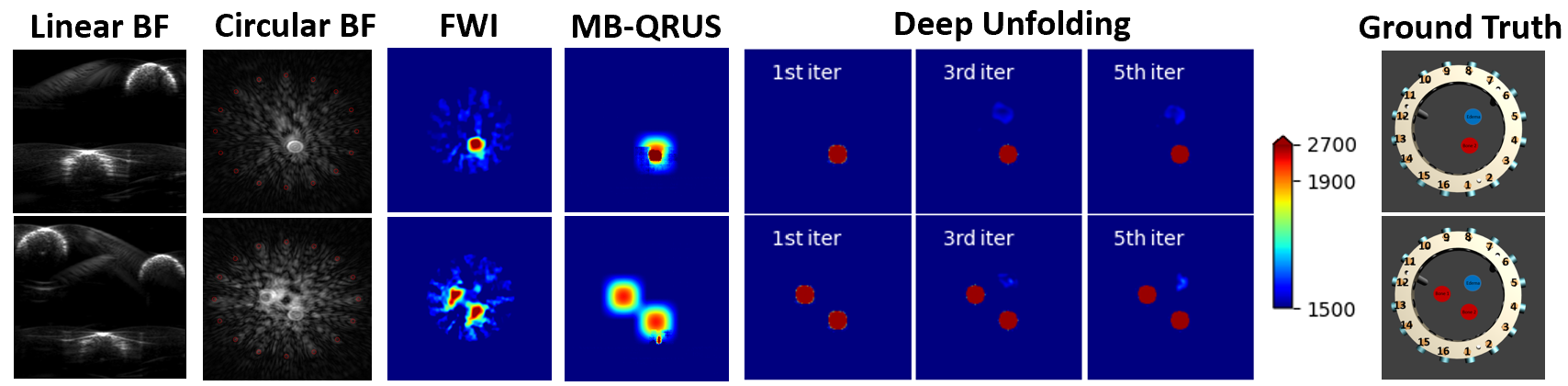}
	}
	% (b) with edema
	\subfigure[\label{fig:res_hardware_2}]{
		\includegraphics[width=0.21\textwidth,keepaspectratio]{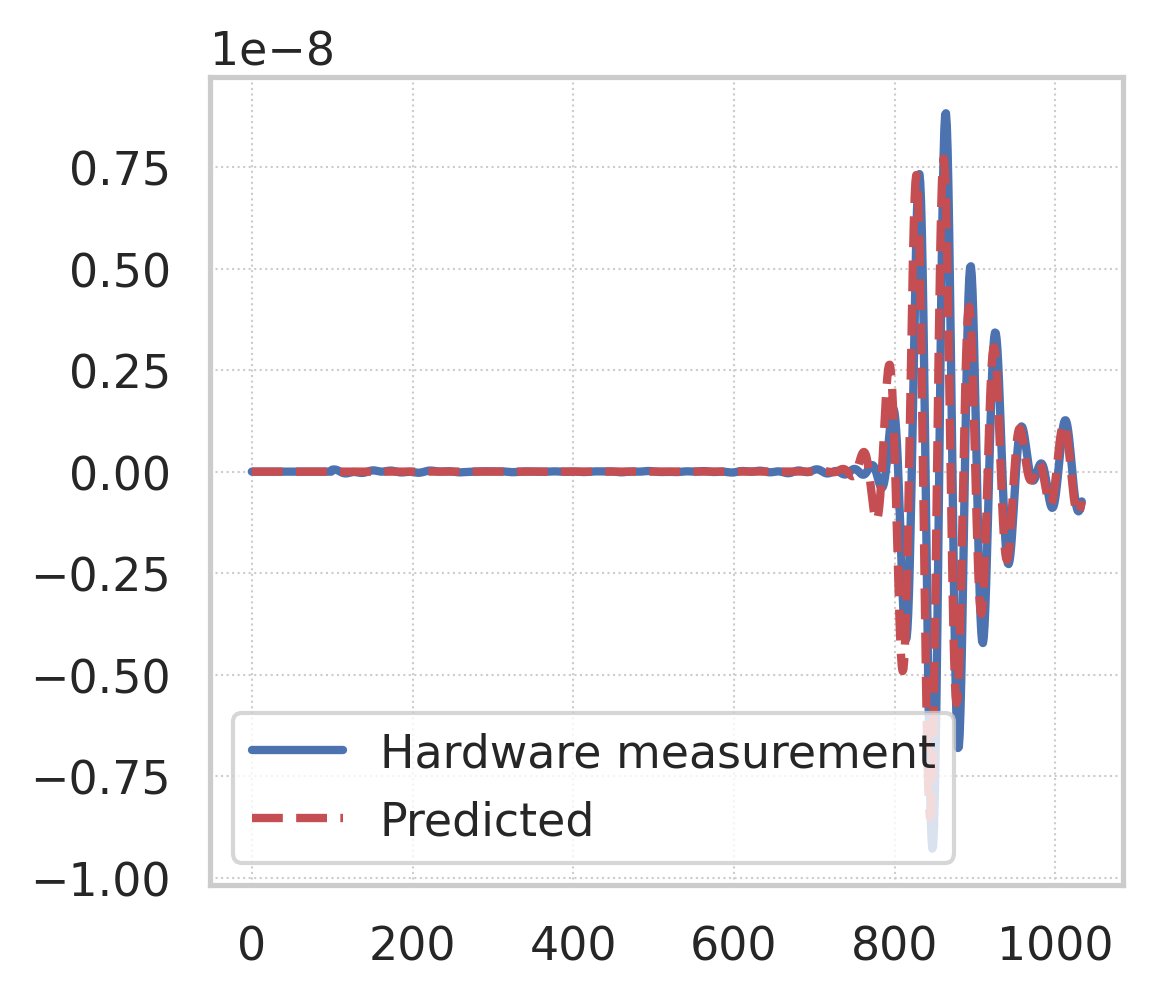}
	}
	\vspace*{-0.8em}
	\caption{Hardware results: (a) Reconstructed SoS maps from the measured phantom data.  “Bone” rods correspond to high SoS value ($\approx$2700 m/s) and “edema” rods correspond to low SoS value ($\approx$1588 m/s). (b) Comparison of hardware measurement and DUFWI-based simulated CD.}
	\label{fig:Compare_all1}
	\vspace*{-1.0em}
\end{figure*}

\vspace{-0.26cm}
\section{Hardware Results}\label{hardware res}
\vspace*{-0.2em}

In this section, we further evaluate DUFWI for practical edema diagnosis using the hardware system described in Section~\ref{sec:harware system}.
To validate the proposed DUFWI method under realistic conditions using our hardware system, we created a circular phantom dataset. This dataset consists of cylindrical rods immersed in a water bath, each containing 1 to 3 rods (diameter: 1 cm) representing bone (2700 m/s) or edema (1588 m/s). These rods were embedded within a homogeneous water background (approximately 1500 m/s) to create strong acoustic contrast, simulating clinically relevant scenarios. GT rod locations are approximate due to manual placement. During training, we use simulated rods with well-defined boundaries to match the experimental setup, while for testing, we employ real maps to assess the generalization capability of DUFWI. This approach introduces a domain gap between training and testing, challenging the model to accurately reconstruct SoS maps from noisy, real-world data.

We generate 16000 simulated data to train the proposed DUFWI architecture. As described earlier, the arm dataset includes both edema and non-edema cases: the non-edema subset models two bones within a homogeneous soft-tissue background, whereas the edema subset adds an elliptical, low speed region to mimic interstitial fluid accumulation. This design supports anatomically realistic training and evaluation.

Fig.~\ref{fig:res_hardware_1} illustrates qualitative reconstructions from measured phantom CD of two test phantoms described in~\ref{sec:harware system} using linear beamforming from the side (by a different probe), circular beamforming (from the circular CD), the conventional FWI, MB-QRUS, and the proposed DUFWI. Linear beamforming yields broad arc-shaped responses and strong multipath clutter, and returns only an intensity envelope without quantitative estimation such as SoS value. Circular beamforming benefits from a larger effective aperture yet still shows pronounced sidelobes and speckle, moreover, the lack of quantitative mapping and the residual artifacts make edema identification unreliable.
%, with the edematous inclusion appearing smeared or confounded by background fluctuations. 
In contrast, DUFWI rapidly forms compact and high-contrast spots at the true locations: by the 1st-3rd iterations the targets are clearly localized, and by the 5th iteration the reconstructions are stable with low background energy and minimal ringing.
MB-QRUS can only provide a coarse localization of the bone structures and completely fails to identify the edema regions. Conventional FWI produces broader artifacts with blurred peaks and occasional position bias. In addition, FWI used 200 iterations, whereas DUFWI used only 5 irerations. The average runtime was $60 \pm 20$ minutes for FWI and $15.83 \pm 3$ seconds for DUFWI, corresponding to an approximate $227\times$ speedup in favor of DUFWI.
% DUFWI enables high-quality ultrasound reconstructions with a reduced number of iterations, providing real-time quantitative recovery of edema morphology and supporting clinical decisions.
% that may mitigate the risk of lymphatic obstruction and treatment-related complications.
%	Totally, DUFWI achieves faster convergence, higher spatial resolution, and quantitatively consistent SoS estimates in the hardware experiments. 
Quantitative results comparing DUFWI and FWI are summarized in Table~\ref{tab:hardware}.
\begin{table}[htbp]
	\vspace{-1.26em}
	\centering
	\caption{Quantitative comparison across all hardware measured data.}
	\vspace{-0.66em}
	\label{tab:hardware}
	\resizebox{\columnwidth}{!}{
		\begin{tabular}{|l|c|c|c|c|c|c|c|}
			\hline
			\multicolumn{8}{|c|}{\textbf{Measured Data}} \\
			\hline
			\multirow{2}{*}{\textbf{Metric}} &
			\multirow{2}{*}{\textbf{FWI}} &
			\multirow{2}{*}{\textbf{MB-QRUS}} &
			\multicolumn{5}{c|}{\textbf{DUFWI}} \\
			\cline{4-8}
			& & & \textbf{1st iter} & \textbf{2nd iter} & \textbf{3rd iter} & \textbf{4th iter} & \textbf{5th iter} \\
			\hline
			SSIM~(\(\uparrow\))& 0.9116 & 0.9195 & 0.9814 & 0.9817 & 0.9834 & 0.9840 & \textbf{0.9846}  \\
			PSNR (dB)~(\(\uparrow\))& 28.41 & 28.80 & 35.04 & 35.54 & 35.63 & 35.65 & \textbf{35.66}  \\
			NMSE (dB)~(\(\downarrow\)) & -23.65 & -27.98 & -30.17 & -30.65 & -30.76 & -30.79 & \textbf{-30.81}  \\
			\hline
		\end{tabular}
	}
	\vspace{-1.5em}
\end{table}

%\begin{table}[htbp]
%	\vspace{-0.5em}
%	\centering
%	\caption{Quantitative comparison across all hardware measured data.}
%	\vspace{-0.5em}
%	\label{tab:hardware}
%	\resizebox{\columnwidth}{!}{
%		\begin{tabular}{|l|c|c|c|c|c|c|}
%			\hline
%			\multicolumn{7}{|c|}{\textbf{Measured Data}} \\
%			\hline
%			\multirow{2}{*}{\textbf{Metric}} & 
%			\multirow{2}{*}{\textbf{\makecell{FWI}}}
%			\multicolumn{5}{c|}{\textbf{DUFWI}} &
%			\\
%			\cline{3-7}
%			& \textbf{1st iter} & \textbf{2nd iter} & \textbf{3rd iter} & \textbf{4th iter} & \textbf{5th iter} & \\
%			\hline
%			SSIM~(\(\uparrow\))& 0.9116 & 0.9814 & 0.9817 & 0.9834 & 0.9840 & \textbf{0.9846}  \\
%			PSNR (dB)~(\(\uparrow\))& 28.41 & 35.04 & 35.54 & 35.63 & 35.65 & \textbf{35.66}  \\
%			NMSE (dB)~(\(\downarrow\)) & -23.65 & -30.17 & -30.65 & -30.76 & -30.79 & \textbf{-30.81}  \\
%			\hline
%		\end{tabular}
%	}
%	\vspace{-0.5em}
%\end{table}
As a representative example from the second scenario in Fig.~\ref{fig:res_hardware_1}, Fig.~\ref{fig:res_hardware_2} compares the hardware–measured CD with the simulated CD from the DUFWI-reconstructed SoS. Prior to the first arrival, both CD remain near the noise. At the onset (around sample index 800), the simulated received signal aligns in time with the measurement. The carrier frequency, envelope shape, and peak amplitude are closely matched, indicating high data consistency between the DUFWI-based results and the physical acquisition. 
%Overall, the strong agreement validates that the DUFWI-reconstructed SoS supports forward simulations that faithfully reproduce measured CD.

Overall, DUFWI enables high-quality US reconstructions with a reduced number of iterations, providing real-time quantitative recovery of edema morphology and supporting clinical decisions. 
%And the simulated forward process within the DUFWI matches the hardware measurement.
%\begin{figure}[htbp]
%	\centering
%	\includegraphics[width=0.33\textwidth, keepaspectratio]{Fig/cd_hardware.png}
%	\caption{Comparison of hardware measurement and DUFWI-based simulated channel data.}  
%	\label{fig:CD_hardware}
%\end{figure}

\vspace{-0.36cm}
\section{Conclusion} \label{sec: conclusion} 
\vspace*{-0.2em}

The proposed DUFWI framework offers a real‐time solution for edema diagnosis.
% by embedding the FWI method into the network architecture. 
By unrolling a FWI algorithm and replacing each gradient step with a learned correction network, DUFWI reconstructs high-fidelity SoS maps from CD.
In extensive 2-D numerical experiments on simulated MNIST, simulated arm datasets, together with measurements on phantom data, DUFWI surpasses classical FWI and the MB-QRUS in reconstruction accuracy. This stems from the model‐driven design, which leverages trainable gradient updates to achieve superior efficiency with less computational overhead.

Looking ahead, within the same DUFWI framework, we can recover additional physical properties, such as density and attenuation, by using forward models whose propagation physics explicitly account for these quantities. Future work will validate these extensions on complex, realistic phantom measurements to expand clinical relevance in brain, breast imaging and other clinical settings.

%Looking ahead, DUFWI can be used to recover additional physical properties by swapping in the appropriate forward model, and to other medical tissue imaging via adjusted propagation models, such as brain and breast imaging. Future work will validate our approach on complex real phantom measurements to broaden its clinical impact.
%and in vivo patient data, particularly for applications like fatty-liver assessment, 
%and explore advanced transmission setups (e.g., linear probes) 
%to broaden its clinical impact.

\end{document}